\documentclass[preprint,12pt]{elsarticle}
\usepackage{hyperref}
\usepackage{color}
\usepackage{graphicx} 
\usepackage{booktabs} 
\usepackage{caption}
\usepackage{subcaption}
\usepackage{amsmath,amssymb,latexsym}
\usepackage{amsfonts}
\usepackage{mathtools}
\usepackage{algorithm}
\usepackage{hyperref}
\usepackage{cleveref}
\usepackage{mathtools}
\usepackage{mathrsfs,bm}
\usepackage{algorithmic}
\usepackage{comment}
\usepackage{dsfont}
\usepackage{amsthm}
\usepackage{siunitx}
\theoremstyle{definition}
\newtheorem{definition}{Definition}

\newtheorem{remark}{Remark}%
\journal{Applied Numerical Mathematics}

\newcommand{\defined}{=}

\begin{document}
\begin{frontmatter}
\title{Bayesian decomposition using Besov priors}

\author[label1]{Andreas Horst}
\ead{ahor@dtu.dk}
\author[label2]{Babak M.  Afkham}
\author[label1]{Yiqiu Dong}
\author[label1]{Jakob Lemvig}
\affiliation[label1]{organization={Department of applied mathematics and computer science, Technical university of Denmark},
            addressline={Matematiktorvet}, 
            city={Kgs. Lyngby},
            postcode={2800}, 
            country={Denmark}}
\affiliation[label2]{organization={Research Unit of Mathematical Sciences, University of Oulu},
            addressline={Pentti Kaiteran katu 1}, 
            city={Linnanmaa}, 
            country={Finland}}
\begin{abstract}
    In many inverse problems, the unknown is composed of multiple components with different regularities, for example, in imaging problems, where the unknown can have both rough and smooth features. We investigate linear Bayesian inverse problems, where the unknown consists of two components: one smooth and one piecewise constant. We model the unknown as a sum of two components and assign individual priors on each component to impose the assumed behavior. We propose and compare two prior models: (i) a combination of a Haar wavelet-based Besov prior and a smoothing Besov prior, and (ii) a hierarchical Gaussian prior on the gradient coupled with a smoothing Besov prior. To achieve a balanced reconstruction, we place hyperpriors on the prior parameters and jointly infer both the components and the hyperparameters. We propose Gibbs sampling schemes for posterior inference in both prior models. We demonstrate the capabilities of our approach on 1D and 2D deconvolution problems, where the unknown consists of smooth parts with jumps. The numerical results indicate that our methods improve the reconstruction quality compared to single-prior approaches and that the prior parameters can be successfully estimated to yield a balanced decomposition.
\end{abstract}

\end{frontmatter}
\newpage

\section{Introduction}
\label{sec:introduction}
Inverse problems concern the estimation of unknown parameters of a system given indirect measurements. Such problems are typically ill-posed, meaning the computation of the solution is heavily sensitive to measurement limitations, measurement noise, and imprecise models. Stable solutions are often obtained through regularization techniques, which imposes penalties on the solution, e.g., periodicity, smoothness, nonsmoothness, or sparsity with respect to an expansion representation.
In many inverse imaging problems the unknown parameters naturally consist of multiple components, e.g., image decomposition of cartoon and texture parts; see, for example, \cite{Chambolle1997,Starck2005,GAO2019,Zhang2021,SHI2024}. Taking the multiple components into account when modeling such inverse problems enables the use of a specific regularizer on each component to achieve a balanced solution that avoids overfitting to a subset of the components.

In this paper, we consider linear inverse problems with two components
\begin{equation}
    \mathbf{y}=\mathbf{A}\left(\mathbf{g}+\mathbf{h}\right)+\epsilon, \quad \mathbf{f}=\mathbf{g}+\mathbf{h}
    \label{eq:InverseProblem}
\end{equation}
where $\mathbf{y}\in \mathbb{R}^{m}$ is the data, $\mathbf{f}\in \mathbb{R}^{n}$ is the unknown which is decomposed into the two components $\mathbf{g},\mathbf{h}\in \mathbb{R}^{n}$, $\mathbf{A}\in \mathbb{R}^{m\times n}$ is the linear forward operator, and $\mathbf{\epsilon}\in \mathbb{R}^{m}$ is additive noise. We assume that the component $\mathbf{g}$ is piecewise constant and the component $\mathbf{h}$ is smooth. These assumptions are reasonable in imaging applications such as anomaly detection in image deblurring~\cite{Kaipio2002,Calvetti2010}, atmospheric emission tomography~\cite{Chung2024}, magnetoencephalography~\cite{Calvetti2011}, and X-ray computed tomography~\cite{Christensen_2024}.

Many variational methods have been proposed for signal or image decomposition \cite{Kutyniok2012,Bobin2007,Gribonval2003,Liu2019}, but all of these methods lack the ability to quantify uncertainties in the decomposition results with respect to noisy data. In this work, we take the Bayesian approach to solve inverse problems as described in \cite{Kaipio2005}, that is, the inverse problem is formulated in a probabilistic sense in terms of random variables and their associated distributions. Regularization is now imposed via the prior probability distribution on the unknown, while model inaccuracies and measurement noise are accounted for in the likelihood function. The solution to a Bayesian inverse problem, called the posterior distribution, is the conditional distribution of the unknown given an observation of the data.
Rather than estimating the full signal $\mathbf{f}$ directly, we seek to reconstruct the individual components $\mathbf{g}$ and $\mathbf{h}$ separately. This objective requires characterizing the joint posterior distribution of the components. We model $\mathbf{g}$ and $\mathbf{h}$ as mutually independent to reflect their distinct structural behavior; for instance, the global smoothness of $\mathbf{h}$ provides no predictive insight into the piecewise transitions of $\mathbf{g}$ prior to observing $\mathbf{y}$. Working in the joint parameter space $(\mathbf{g}, \mathbf{h})$ under this independence assumption facilitates the separate reconstruction of each component. This formulation allows us to explicitly impose distinct prior information on each component, which is otherwise difficult to enforce separately when considering the posterior distribution of the full signal $\mathbf{f}$.
Under the mutual independence assumption between the components $\mathbf{g}$, $\mathbf{h}$ and the noise $\mathbf{\epsilon}$, the joint posterior distribution associated with the inverse problem \eqref{eq:InverseProblem} is given by Bayes' theorem \cite{Kaipio2005} as
\begin{equation*}
    \pi_{\text{post}}(\mathbf{g},\mathbf{h}|\mathbf{y}=\mathbf{y}^{*})\propto \pi_{\text{like}}(\mathbf{y}=\mathbf{y}^{*}|\mathbf{g},\mathbf{h})\pi_{\text{prior}}(\mathbf{g})\pi_{\text{prior}}(\mathbf{h}).
    \label{eq:Posterior}
\end{equation*}
Here $\mathbf{y}^{*}\in \mathbb{R}^{m}$ is the observed data, $\pi_{\text{like}}$ is the likelihood function, $\pi_{\text{prior}}$ is the prior distribution and $\pi_{\text{post}}$ is the posterior distribution.
Characterizing the posterior does not only give statistical estimates of the unknown, it also provides the related uncertainties in contrary to standard regularization techniques.

The formulation of prior distributions in Bayesian inverse problems is crucial for obtaining accurate estimates of the unknown and ensuring that its assumed properties are properly enforced. In this work, we focus on two classes of priors: smoothness priors for representing the component $\mathbf{h}$, and priors that promote the piecewise constant structure for representing the component $\mathbf{g}$. One way of modeling smoothness priors is to construct prior distributions that impose smoothness in a specific function space. These models include Gaussian priors that are realized in Sobolev spaces \cite{Dashti2017,Henderson2024} and Besov priors that are realized in Besov spaces \cite{Lassas2009,Dashti&Stuart2012}. These models offer a consistent representation of smoothness and provide explicit control over it through the choice of prior parameters. The modeling of priors that promote piecewise constant structures is an active area of research, with several approaches available such as heavy-tailed priors \cite{suuronen2022cauchy,suuronen2023bayesian,uribe2022hybrid,Senchukova_2024,Flock2024}, hierarchical shrinkage priors \cite{calvetti2019hierachical,uribe2023horseshoe,Calvetti2020,Dong_2023,Waniorek_2023}, machine-learning priors \cite{li2022bayesian,Glaubitz2024,Laumont2022}, and Haar-wavelet-based Besov priors \cite{kekkonen2023random,Siltanen2013,Kolehmainen_2012,Horst2024}. Each of these methods has its strengths and limitations and should be selected based on the specific constraints of the problem at hand. We focus primarily on Besov priors as they offer a flexible framework for modeling both smooth and piecewise constant structures, while also allowing efficient computation and interpretable parameter choices.

In addition, Besov spaces provide the theoretical foundation for this broad class of functions, characterizing both smooth regions and localized singularities. Indeed, functions combining smooth variations with discontinuities can be recovered in a minimax-optimal sense in the white-noise model using nonlinear wavelet thresholding \cite{Donoho1994Ideal,Donoho1995Adapting,Donoho1998Minimax}, Bayesian wavelet thresholding \cite{Abramovich2002Wavelet}, or hierarchical Bayesian formulations \cite{Agapiou2024Adaptive,Giordano2023Besov}. While these results demonstrate the theoretical properties of the aforementioned Besov-based approaches, our focus is on applications in linear inverse problems under a decomposition model. By modeling each component according to its structure, decomposition approaches offer the flexibility to reconstruct different features separately with uncertainty quantification.

The selection of parameters, which controls the relative contribution of each component's prior information, is crucial for obtaining a balanced decomposition under the two-component model. In classical regularization methods, these parameters are often tuned manually based on visual quality, and there has been limited work on automatic parameter selection for decomposition (see, e.g., \cite{Chung2024}). However, in the Bayesian framework, we can include these parameters as hyperparameters and estimate them jointly with the decomposition solution via the posterior. The main challenges lie in how to design priors and hyperpriors to ensure the components are well separated.

In this paper, we propose two Bayesian methods to solve linear inverse problems, where the unknown consists of two components with different properties.
In the first method, we introduce a novel way of combining two distinct Besov priors by formulating a two-Besov decomposition model. We utilize the Besov prior with a Haar wavelet basis to impose jump discontinuities and the Besov prior with a smooth wavelet to impose smoothness. The No-U-Turn sampler (NUTS)\cite{Hoffman2014} was applied to sample the posterior.
In the second method, we propose a new hierarchical posterior to incorporate the parameters that control the strengths of the priors as hyperparameters. In this method, we use a combination of a hierarchical Gaussian prior on the discrete derivative to impose sparsely located jumps and a Besov prior with a smooth wavelet to impose localized smoothness. To sample this hierarchical posterior, we introduce a Gibbs sampling scheme, where marginal distributions either have closed-form expressions or can be effectively sampled using a Randomize-Then-Optimize (RTO) method \cite{Bardsley2014,Horst2024}. Both methods are tested on 1D and 2D deconvolution problems. The numerical results indicate that the decomposition method yields better reconstructions compared to methods using a single Besov prior. Furthermore, the hierarchical posterior can achieve automatic parameter selection for decomposition problems.

The main contributions of this paper are:
\begin{itemize}
    \item We introduce Bayesian decomposition methods that can jointly recover smooth and piecewise constant components and quantify uncertainties in the reconstruction.
    \item We propose a novel two-Besov decomposition model that incorporates two distinct Besov priors  
          within a unified Bayesian framework.
    \item We propose a new hierarchical posterior to achieve automatic parameter selection in decomposition problems.
    \item We introduce a Gibbs sampling algorithm to sample the posteriors in the proposed methods while estimating the prior strength parameters simultaneously.
    \item We demonstrate the performance of our methods in 1D and 2D deconvolution problems.
\end{itemize}

This paper is structured as follows. In Section~\ref{sec:BesovDecomposition}, we briefly review the concepts of wavelets and Besov priors and introduce a Bayesian decomposition model that combines two different Besov priors. In Section~\ref{sec:hier-decomp-meth}, we present a hierarchical posterior to include prior parameters as hyperparameters for a two-component linear Bayesian inverse problem. Additionally, we derive a Gibbs sampling method for posterior inference and parameter estimation. In Section~\ref{sec:numer-exper}, we illustrate our computational methods on deconvolution problems and showcase the properties of the models addressed. We end the paper with a conclusion in Section~\ref{sec:conclusion}.

\section{Besov decomposition}
\label{sec:BesovDecomposition}
Besov priors are wavelet-based and allow explicit control over regularity in a class of smoothness spaces called Besov spaces. These priors are flexible, and depending on the prior setup, they can be used to impose either piecewise constant or different orders of smoothness.

\subsection{Besov priors and wavelets}
\label{sec:Besovpriors}
Before defining Besov priors, we briefly recall how to construct orthonormal wavelet bases on $L^{2}(\mathbb{R}^{d})$ and $L^{2}(\mathbb{T}^{d})$, where $\mathbb{T}^{d}$ is the $d$-dimensional torus with $d\in \mathbb{N}$.

We first define the combined dilation and translation $f_{j,k}$ of a function $f\in L^{2}(\mathbb{R})$ as
\begin{equation*}
    \label{eq:dilation-translation}
    f_{j,k}(x) = 2^{j/2}f(2^{j}x-k) \quad \text{for a.e. } x\in \mathbb{R},
\end{equation*}
where $j$ and $k$ are integers. We denote the scaled and translated scaling and wavelet functions by $\psi_{j,k,0} = \phi_{j,k}$ and $\psi_{j,k,1} = \psi_{j,k}$. Whenever the collection $\left\{\psi_{0,k,0} \mid k\in \mathbb{Z}\right\}\cup\left\{\psi_{j,k,1} \mid j\in \mathbb{N}_{0},k\in \mathbb{Z}\right\}$ forms an orthonormal basis of $L^{2}(\mathbb{R})$, the function $\psi$ is said to be a wavelet, and $\phi$ is called a scaling function.

Following \cite{Schwab2024}, we let $\phi, \psi \in C^{r}(\mathbb{R})$ with $r\geq 1$ be compactly supported scaling and wavelet functions, respectively. There exist choices of such wavelets $\psi$, where the associated scaling function $\phi$ generates a multiresolution analysis (MRA) of $L^{2}(\mathbb{R})$. An example is the Daubechies wavelet of order $M$, denoted by DB(M)-wavelet in the following, which is supported in $[0,2M-1]$ and belongs to $C^{\gamma M}(\mathbb{R})$ for $\gamma \approx 0.2075$ \cite{Daubechies1988Orthonormal}.

Next, we need to lift the one-variable wavelet and scaling functions to $L^{2}(\mathbb{R}^{d})$.
One way of generating an MRA on $L^{2}(\mathbb{R}^{d})$ is to use a tensor-product of MRAs on $L^{2}(\mathbb{R})$. Define the index sets as $\mathcal{L}_{0}\defined\{0,1\}^{d}$ and $\mathcal{L}_{1}\defined \mathcal{L}_{0}\setminus \{0,0,\ldots,0\}$. Then, for any $\ell \in \mathcal{L}_{0}$, we define the generators as
\begin{equation}
    \psi_{j,k,\ell}(x)\defined \prod_{i=1}^{d} \psi_{j,k_{i},\ell_{i}}(x_{i}),  \quad \text{for a.e. } x = (x_{1},x_{2},\ldots,x_{d}) \in \mathbb{R}^{d},
    \label{eq:Tensorwavelet}
\end{equation}
where $j\in \mathbb{N}_{0}$, $k\in \mathbb{Z}^{d}$ and $\ell = (\ell_{1},\ell_{2},\ldots,\ell_{d})\in \mathcal{L}_{0}$.  Note that $\ell = (0,0,\ldots,0)$ corresponds to the scaling function, while each $\ell \in \mathcal{L}_{1}$ corresponds to the $2d-1$ wavelet functions. The collection $\left\{\psi_{0,k,(0,0,\ldots,0)}\},\; k\in \mathbb{Z}^{d}\right\}$ $\cup \left\{\psi_{j,k,\ell} \mid j\in \mathbb{N}_0,\; k\in \mathbb{Z}^{d},\; \ell \in \mathcal{L}_{1}\right\}$ forms an orthonormal basis of $L^{2}(\mathbb{{R}}^{d})$ \cite{mallat1999wavelet}.

The orthonormal wavelet basis of $L^{2}(\mathbb{R}^{d})$ can be used to create an orthonormal basis of $L^{2}(\mathbb{T}^{d})$ by periodization. Define the index set
$\mathcal{K}_{j}\defined\{k\in \mathbb{Z}^{d} \mid 0\leq k_{1},k_{2},\ldots,k_{d}\leq 2^{j}-1\}$.
Then the 1-periodization of the wavelets in \eqref{eq:Tensorwavelet} is defined, for $j\in \mathbb{N}_{0},k\in \mathcal{K}_{j},\ell\in \mathcal{L}_{0}$, as
\begin{equation*}
    \label{eq:12}
    \psi^{\text{per}}_{j,k,\ell}(x)\defined \sum_{n\in \mathbb{Z}^{d}}\psi_{j,k,\ell}(x-n), \quad \text{for a.e. } x\in \mathbb{T}^{d}.
\end{equation*}
Set the index set as $\mathcal{I}_{\psi}\defined\{(0,0,0,\ldots,0)\}\cup\{j\in \mathbb{N}_{0},k\in\mathcal{K}_{j},\ell \in \mathcal{L}_{1}\}$ then the collection $\{\psi^{\text{per}}_{j,k,\ell},(j,k,\ell)\in \mathcal{I}_{\psi}\}$ is an orthonormal basis of $L^{2}(\mathbb{T}^{d})$ \cite{triebel2008}.
The 1-periodic orthonormal wavelet basis can, under conditions on its smoothness, be used to define a class of smoothness spaces known as Besov spaces \cite{triebel2006}.

\begin{definition}
    \label{def:Besov}
    Let $s>0, p\in [1,\infty)$, and assume that the wavelets belong to $C^{r}(\mathbb{T}^{d})$ with $r > s$. We define the Besov spaces on $\mathbb{T}^{d}$ as the family of spaces
    \begin{equation*}
        \label{eq:Besov-space}
        B_{p,p}^{s}(\mathbb{T}^{d})\coloneqq \left\{f\in L^{p}(\mathbb{T}^{d}) \colon \|f\|_{B_{p,p}^{s}(\mathbb{T}^{d})}<\infty\right\},
    \end{equation*}
    with Besov norms defined by
    \begin{equation*}
        \label{eq:Besovnorm}
        \|f\|_{B_{p,p}^{s}(\mathbb{T}^{d})}\defined \left(\sum_{(j,k,\ell)\in \mathcal{I}_{\psi}}2^{jp(s+\frac{d}{2}-\frac{d}{p})}|w_{j,k,\ell}|^{p}\right)^{1/p}.
    \end{equation*}
    where $w_{j,k,\ell}$ are the wavelet coefficients given by
    \begin{equation*}
        \label{eq:waveletcoeff}
        w_{j,k,\ell}=\int_{\mathbb{T}^{d}}f(x)\overline{\psi^{\text{per}}_{j,k,\ell}(x)} \;dx.
    \end{equation*}
\end{definition}

Besov priors can now be constructed as a wavelet expansion where the expansion coefficients are chosen at random according to generalized Gaussian random variables. For $p\in [1,\infty)$ we define a sequence $\Xi\defined \{\xi_{j,k,\ell},(j,k,\ell)\in \mathcal{I}_{\psi}\}$ of independent identically distributed (i.i.d.) generalized Gaussian random variables, where each $\xi_{j,k,\ell}$ is distributed according to a Lebesgue density given by
\begin{equation}
    \label{eq:p-Gauss-dist}
    \pi_{\xi}(x) \defined \frac{1}{c_{p}}\exp\left(-|x|^{p}\right),\; c_{p}\defined\int_{\mathbb{R}}\exp\left(-|x|^{p}\right)dx,\;  x\in \mathbb{R}.
\end{equation}
To define the random wavelet expansion, we first have to establish the probability space for the coefficient sequence. Let $\mathbb{Q}_{p}$ be the measure associated with a one-dimensional generalized Gaussian random variable on $(\mathbb{R},\mathcal{B}(\mathbb{R}))$. Formally, we take the countable product of these one-dimensional spaces:
\begin{equation*}
    \label{eq:13}
    \Omega \defined \mathbb{R}^{\infty}, \quad \mathcal{B}\defined\bigotimes_{n\in \mathbb{N}}\mathcal{B}(\mathbb{R}), \quad \mathbb{Q}\defined\bigotimes_{n\in \mathbb{N}}\mathbb{Q}_{p}.
\end{equation*}
By Kolmogorov’s extension theorem, the infinite product measure $\mathbb{Q}$ is well-defined on the cylinder $\sigma$-algebra $\mathcal{B}$, giving $(\Omega,\mathcal{B},\mathbb{Q})$ a complete probability-space structure for the sequence $\Xi$.

\begin{definition}
    \label{def:besov-rand-var}
    Let $s>0,p\in[1,\infty)$ and let $\Xi$ be the sequence of i.i.d.~generalized Gaussian random variables. We define a random wavelet expansion as the mapping
    \begin{equation}
        T:\Omega \rightarrow B_{p,p}^{s}(\mathbb{T}^{d}),\;\; \Xi \mapsto \sum_{(j,k,\ell)\in \mathcal{I}_{\psi}}2^{-j(s+\frac{d}{2}-\frac{d}{p})}\xi_{j,k,\ell}(\omega)\psi^{\text{per}}_{j,k,\ell}
        , \;\omega \in \Omega.
        \label{eq:Randomwaveletexpansion}
    \end{equation}
    We say that the random wavelet expansion generates a Besov random variable $T_{p,\psi}^{s}$.
\end{definition}

The random wavelet expansion in \eqref{eq:Randomwaveletexpansion} almost surely takes values in the space $B_{p,p}^{t}(\mathbb{T}^{d})$ if and only if $t<s-\frac{d}{p}$ \cite{Lassas2009} and generates a Besov probability measure defined by a pushforward measure $\mathbb{Q}\circ T^{-1}$ \cite{Dashti&Stuart2012}.
Formally, this gives rise to a Besov prior with Lebesgue density given by
\begin{equation}
    \label{eq:3}
    \pi_{T_{p,\psi}^{s}}(f) \propto \exp(-\|f\|_{B_{p,p}^{s}(\mathbb{T}^{d})}^{p}).
\end{equation}
However, the Lebesgue density in \eqref{eq:3} is not well-defined as a function on $B_{p,p}^{s}(\mathbb{T}^{d})$ since the space is infinite-dimensional. Instead, we interpret the prior as a Radon measure $\mathbb{Q}\circ T^{-1}$ that coincides with the finite-dimensional exponential-power densities on all wavelet cylinder sets.

\subsection{Discrete Besov priors}
\label{sec:discr-besov-priors}

Besov priors can be discretized by truncating the wavelet expansion in \eqref{eq:Randomwaveletexpansion} to the finite index set $\mathcal{I}_{\psi,J}\coloneqq\left\{(0,0,\ldots,0)\right\}\cup \bigl\{j\in\{0,1,\ldots, J-1\},$ $k\in \mathcal{K}_{j},\ell \in \mathcal{L}_{1}\bigr\}$ for some integer $J\geq 1$, which results in a $|\mathcal{I}_{\psi,J}|=2^{Jd}$ coefficient representation of $f$. Set $n=2^{Jd}$ and let $\mathbf{S}\in \mathbb{R}^{n\times n}$ be the Besov scaling matrix which is a diagonal matrix with entries
\begin{equation}
    \label{eq:S_matrix}
    \mathbf{S}_{0,0}= 1, \quad \mathbf{S}_{i,i}=2^{j(s+d/2-d/p)},\; \text{for }2^{jd}+1\leq i \leq 2^{(j+1)d}.
\end{equation}
Let $\mathbf{W}_{\psi}\in \mathbb{R}^{n\times n}$ be a discrete wavelet transform associated to the wavelet $\psi$ in  such a way that
\begin{equation}
    \label{eq:W_matrix}
    \mathbf{W}_{\psi}\mathbf{f}\approx \{\langle f,\psi_{j,k,\ell}^{\text{per}}\rangle_{L^{2}(\mathbb{T}^{d})}\}_{(j,k,\ell)\in\mathcal{I}_{\psi,J}}.
\end{equation}
We will use the fast wavelet transform \cite{Mallat1989} implemented in the Pywavelets module \cite{Lee2019} for computing the wavelet transform with $\mathbf{W}_{\psi}$.
The continuous Besov norm can be approximated by a discrete version as
\begin{equation*}
    \label{eq:6}
    \|f\|_{B_{p,p}^{s}(\mathbb{T}^{d})}\approx \|\mathbf{S}\mathbf{W}_{\psi}\mathbf{f}\|_{p}.
\end{equation*}
The discrete Besov prior distribution is then given by
\begin{equation*}
    \label{eq:7}
    \pi_{T_{p,\psi}^{s}}(\mathbf{f})\propto \exp\left(-\lambda\|\mathbf{S}\mathbf{W}_{\psi}\mathbf{f}\|_{p}^{p}\right),
\end{equation*}
where $\lambda>0$ is the prior strength parameter.

\subsection{The posterior for Bayesian decomposition}
\label{sec:post-bayes-decomp}
We consider a decomposition model in which the unknown $f$ is split into two components, $f = \mathbf{g} + \mathbf{h}$, with different structural properties. To impose piecewise constant characteristics, we let $\mathbf{g}$ follow a Besov prior $\mathbf{g}\sim T_{p_{g},\psi_{g}}^{s_{g}}$ where $\psi_{g}$ is the Haar wavelet. To impose smoothness in $\mathbf{h}$, we assign a Besov prior $\mathbf{h} \sim T_{p_h, \psi_h}^{s_h}$, with $\psi_{h}$ a smooth wavelet, e.g., a high-order Daubechies wavelet.

Assuming prior independence between the components $\mathbf{g}$ and $\mathbf{h}$ to reflect their distinct regularities, the joint prior factorizes as a product measure of the form
\begin{equation*}
    \pi_{\text{prior}}(\mathbf{g})\pi_{\text{prior}}(\mathbf{h}) \propto \exp\left(-\lambda_{g}\|\mathbf{S}_{g}\mathbf{W}_{\psi_{g}}\mathbf{g}\|_{p_{g}}^{p_{g}}-\lambda_{h}\|\mathbf{S}_{h}\mathbf{W}_{\psi_{h}}\mathbf{h}\|_{p_{h}}^{p_{h}}\right),
    \label{eq:JointBesovPrior}
\end{equation*}
where $\lambda_{g}, \lambda_{h}>0$ are the prior strength parameters associated to the components. This formulation allows us to combine different Besov priors in one posterior.

In the inverse problem defined in \eqref{eq:InverseProblem}, we assume the noise is Gaussian  $\epsilon \sim N(\mathbf{0},\sigma^{2}\mathbf{I}_m)$ with zero mean and covariance matrix $\sigma^{2}\mathbf{I}_m$, where $\sigma>0$ and $\mathbf{I}_m\in \mathbb{R}^{m\times m}$  is the identity matrix. This assumption leads to a likelihood function for the components given by
\begin{equation}
    \pi_{\text{like}}(\mathbf{y}=\mathbf{y}^{*}|\mathbf{g},\mathbf{h})\propto \exp\left(-\frac{1}{2\sigma^{2}}\|\mathbf{A}(\mathbf{g}+\mathbf{h})-\mathbf{y}^{*}\|_{2}^{2}\right).
    \label{eq:Likelihood}
\end{equation}

The solution to the Bayesian decomposition problem is given by the posterior distribution, which combines the prior and likelihood. According to Bayes' theorem~\cite{Kaipio2005}, the posterior is proportional to the product of the likelihood and the prior, i.e.,

\begin{align}
    \pi_{\text{post}}(\mathbf{g},\mathbf{h}|\mathbf{y}=\mathbf{y}^{*})\propto \exp\Bigl( & -\frac{1}{2\sigma^{2}}\|\mathbf{A}(\mathbf{g}+\mathbf{h})-\mathbf{y}^{*}\|_{2}^{2} \nonumber                                                                      \\
                                                                                         & -\lambda_{g}\|\mathbf{S}_{g}\mathbf{W}_{\psi_{g}}\mathbf{g}\|_{p_{g}}^{p_{g}}-\lambda_{h}\|\mathbf{S}_{h}\mathbf{W}_{\psi_{h}}\mathbf{h}\|_{p_{h}}^{p_{h}}\Bigr).
    \label{eq:Posterior_1}
\end{align}
When the prior strength parameters $\lambda_{g}$ and $\lambda_{h}$ are fixed, we use the No-U-Turn sampler (NUTS)~\cite{Hoffman2014} to sample from the posterior distribution in \eqref{eq:Posterior_1}. NUTS is a Hamiltonian Markov chain Monte Carlo method with automatic adaptive step size control, which usually makes it efficient and eliminates the need for manual tuning of the sampler. The negative log posterior of \eqref{eq:Posterior_1} is almost everywhere differentiable for all choices of prior parameters $s_{g}, s_{h}>0$ and $p_{g}, p_{h}\in [1,\infty)$, which makes NUTS applicable in this setting.

In decomposition problems, the results strongly depend on the strength parameters $\lambda_g$ and $\lambda_h$, which define the balance between two components. The choice of the strength parameters is crucial and challenging. Under Bayesian statistics, we can introduce a hierarchical model to incorporate both parameters as unknown random variables, called hyperparameters, in the posterior\footnote{The model is derived in \ref{sec:gibbs-sampl-hier}. For numerical experiments, see Figs.~\ref{fig:Decomp_deconv_besov_param} and~\ref{fig:Decomp_deconv_besov_x}.}. Unfortunately, even when the two priors are constructed from structurally different wavelets, the resulting hierarchical model suffers
from an identifiability problem. In general, two wavelet bases have high mutual coherence defined as the maximum absolute value of the inner product between two wavelet functions from different bases \cite{MR1963681}. In particular, we see a large coherence between scaling functions and other low-frequency wavelets from the different bases. This high mutual coherence leads to a lack of identifiability in the posterior distribution. When the strength parameters are considered as random variables, it becomes difficult to distinguish the contributions between two Besov priors, as they can represent similar features in the unknown. This lack of identifiability can result in a multimodal posterior, as different parameter choices may cause different priors to dominate the reconstruction. Sampling from multimodal posteriors is a well-known challenge for Gibbs sampling and Markov chain Monte Carlo methods in general, and addressing it is beyond the scope of this paper. The identifiability issues are numerically illustrated in Section~\ref{sec:numer-exper}.

To address this issue, we replace the Besov prior for the piecewise constant component with a hierarchical Gaussian prior on the discrete gradient. The hierarchical Gaussian prior, through its sparsity-promoting hyperprior, shrinks most discrete gradients towards zero but allows occasional jumps. As a result, even though the gradient and wavelet systems are not strictly incoherent, the strength of the sparsity enforces a clearer distinction between the two representations.

\section{Hierarchical Gaussian-Besov decomposition method}
\label{sec:hier-decomp-meth}

In this section, we propose an alternative prior for the piecewise constant component $\mathbf{g}$ by modeling its discrete gradient as a Gaussian random vector with a diagonal covariance matrix. The associated variance parameters are treated as hyperparameters.  Combined with a smooth Besov prior for $\mathbf h$, this yields a new hierarchical Bayesian decomposition model that can be sampled efficiently with Gibbs updates and that avoids the identifiability issues of the two-Besov formulation.

Assume that the gradient of $\mathbf{g}$, denoted $\mathbf{D}_{d}\mathbf{g}\in\mathbb{R}^{dn}$, follows a Gaussian distribution $\mathcal{N}(\mathbf{0},\mathbf{\Lambda})$ with mean $\mathbf{0}$ and covariance $\mathbf{\Lambda}$. The 1D gradient operator is defined using a forward finite difference scheme with periodic boundary conditions as
\begin{equation*}
    \label{eq:grad1d}
    \mathbf{D}_{1}=\begin{bmatrix}
        -1 & 1  &        &        &    \\
           & -1 & 1      &        &    \\
           &    & \ddots & \ddots &    \\
           &    &        & -1     & 1  \\
        1  &    &        &        & -1
    \end{bmatrix}\in\mathbb{R}^{n\times n}.
\end{equation*}
The matrix representation of the higher-dimensional gradient operator can be generated by $\mathbf{D}_1$ and the Kronecker product, e.g., the 2D gradient operator is given as
\begin{equation*}
    \label{eq:grad2d}
    \mathbf{D}_{2}=\begin{bmatrix}
        \mathbf{I}_n \otimes \mathbf{D}_{1} \\ \mathbf{D}_{1}\otimes \mathbf{I}_n
    \end{bmatrix}\in\mathbb{R}^{2n\times n},
\end{equation*}
where $\otimes$ denotes the Kronecker product. Referring to the work in \cite{Calvetti2020}, we consider a diagonal covariance $\mathbf{\Lambda}$ and assume that the diagonal entries are i.i.d.~random variables following the inverse gamma distribution $\mathcal{IG}(\alpha_{1},\beta_{1})$. This hierarchical Gaussian prior imposes sparsity on the gradient of $\mathbf{g}$. For the smooth component $\mathbf{h}$, we continue to use a discrete Besov prior $\mathbf{h}\sim T_{p_h,\psi_{h}}^{s_{h}}$ with $p_h=2$ and a smooth wavelet $\psi_{h}$. Furthermore, we set the prior strength parameter $\lambda_{h}$ as a hyperparameter following the gamma distribution $\mathcal{G}(\alpha_{2},\beta_{2})$, which is conjugate to the Gaussian distribution. Combining with the Gaussian likelihood, we form the joint posterior of the components and hyperparameters as
\begin{align}
    \pi_{\text{post}}(\mathbf{g},\mathbf{h},\mathbf{\Lambda},\lambda_{h}|\mathbf{y}=\mathbf{y}^{*})
     & \propto \pi_{\text{like}}(\mathbf{y}=\mathbf{y}^{*}|\mathbf{g},\mathbf{h})\pi_{\text{prior}}(\mathbf{g}|\mathbf{\Lambda})\pi_{\text{prior}}(\mathbf{h}|\lambda_{h}) \pi_{\text{hyp}}(\mathbf{\Lambda}) \pi_{\text{hyp}}(\lambda_{h}) \nonumber \\
     & \propto \lambda_{h}^{n/2+\alpha_{2}-1}\left(\prod_{i=1}^{dn}\mathbf{\Lambda}_{i,i}\right)^{-\alpha_{1}-3/2} \nonumber                                                                                                                          \\
     & \phantom{\propto \,\,}\exp\Bigl(-\frac{1}{2\sigma^{2}}\|\mathbf{A}(\mathbf{g}+\mathbf{h})-\mathbf{y}^{*}\|_{2}^{2}-\frac{1}{2}\|\mathbf{\Lambda}^{-1/2}\mathbf{D}_{d}\mathbf{g}\|_{2}^{2}\nonumber                                             \\
     & \phantom{\propto \,\, \exp \,} -\frac{\lambda_{h}}{2}\|\mathbf{S}_{h}\mathbf{W}_{\psi_{h}}\mathbf{h}\|_{2}^{2}-\beta_{2}\lambda_{h}-\beta_{1}\sum_{i=1}^{dn}\mathbf{\Lambda}_{i,i}^{-1}\Bigr). \label{eq:Posterior_2}
\end{align}
To sample from the posterior in \eqref{eq:Posterior_2}, we use a Gibbs sampling scheme outlined in Algorithm~\ref{alg:Gibbs_Scheme}, which samples components and hyperparameters sequentially from their marginal distributions given as follows.

    \begin{align*}
        \pi_{1}(\mathbf{g}|\mathbf{y}^{*},\mathbf{h},\mathbf{\Lambda}) & \propto \exp\left(-\frac{1}{2\sigma^{2}}\|\mathbf{A}(\mathbf{g}+\mathbf{h})-\mathbf{y}^{*}\|^{2}_{2}-\frac{1}{2}\|\mathbf{\Lambda}^{-1/2}\mathbf{D}_{d}\mathbf{g}\|_{2}^{2}\right),                                       \\
        \pi_{2}(\mathbf{h}|\mathbf{y}^{*},\mathbf{g},\lambda_{h})      & \propto \exp\left(-\frac{1}{2\sigma^{2}}\|\mathbf{A}(\mathbf{g}+\mathbf{h})-\mathbf{y}^{*}\|_{2}^{2}-\frac{\lambda_h}{2}\|\mathbf{S}_{h}\mathbf{W}_{\psi_{h}}\mathbf{h}\|_{2}^{2}\right),                                 \\
        \pi_{3}(\mathbf{\Lambda}|\mathbf{g})                           & \propto \left(\prod_{i=1}^{dn}\mathbf{\Lambda}_{i,i}\right)^{-\alpha_{1}-3/2}\exp\left(-\frac{1}{2}\|\mathbf{\Lambda}^{-1/2}\mathbf{D}_{d}\mathbf{g}\|_{2}^{2}-\beta_{1}\sum_{i=1}^{dn}\mathbf{\Lambda}_{i,i}^{-1}\right), \\
        \pi_{4}(\lambda_{h}|\mathbf{h})                                & \propto \left(\lambda_{h}\right)^{n/2+\alpha_{2}-1}\exp\left(-\frac{\lambda_h}{2}\|\mathbf{S}_{h}\mathbf{W}_{\psi_{h}}\mathbf{h}\|_{2}^{2}-\beta_{2}\lambda_{h}\right). 
    \end{align*}

To sample the marginal distributions $\pi_{1}$ and $\pi_{2}$, we notice that they are Gaussian and can be sampled efficiently using the Randomize-Then-Optimize (RTO) method \cite{Bardsley2014,Horst2024}. RTO draws samples from a Gaussian distribution by solving the following randomly perturbed linear least squares problems
\begin{subequations}
    \begin{align}
         & \arg \min_{\substack{\mathbf{g}}}\frac{1}{2}\left\|
        \begin{bmatrix}
            \frac{1}{\sigma}\mathbf{A} \\
            \mathbf{\Lambda}^{-1/2}\mathbf{D}_d
        \end{bmatrix}\mathbf{g}-\begin{bmatrix}
                                    \frac{1}{\sigma}\mathbf{y}^{*}-\frac{1}{\sigma}\mathbf{A}\mathbf{h} \\
                                    \mathbf{0}
                                \end{bmatrix}
        - \bm{\eta}^{*}\right\|_{2}^{2},\quad \bm{\eta}\sim \mathcal{N}(\mathbf{0},\mathbf{I}_{m+dn}), \label{eq:g_samp} \\
         & \arg \min_{\substack{\mathbf{h}}}\frac{1}{2}\left\|
        \begin{bmatrix}
            \frac{1}{\sigma}\mathbf{A} \\
            \sqrt{\lambda_{h}}\mathbf{S}_{h}\mathbf{W}_{\psi_{h}}
        \end{bmatrix}
        \mathbf{h}-\begin{bmatrix}
                       \frac{1}{\sigma}\mathbf{y}^{*}-\frac{1}{\sigma}\mathbf{A}\mathbf{g} \\
                       \mathbf{0}
                   \end{bmatrix}
        -\bm{\zeta}^{*}\right\|_{2}^{2}, \quad \bm{\zeta}\sim N(\mathbf{0},\mathbf{I}_{m+n}), \label{eq:h_samp}
    \end{align}
\end{subequations}
where $\bm{\eta}^{*}$ and $\bm{\zeta}^{*}$ are realizations of $\bm{\eta}$ and $\bm{\zeta}$, respectively. We solve the linear least squares problems using a CGLS algorithm \cite{hestenes1952methods}. To sample the covariance matrix $\mathbf{\Lambda}$ we notice that $\pi_{3}$ is an independent product of
\begin{equation}
    \label{eq:Hyperparameters_1}
    \mathbf{\Lambda}_{i,i}\sim \mathcal{IG}\left(\alpha_{1}+1/2,\beta_{1}+\left(\mathbf{D}_{d}\mathbf{g}\right)^{2}_{i}\right),
\end{equation}
such that we can sample each element independently and directly from an inverse gamma distribution. Finally, $\pi_{4}$ is a gamma distribution
\begin{equation}
    \label{eq:Hyperparameters_2}
    \lambda_{h}\sim \mathcal{G}\left(n/2+\alpha_{2},\beta_{2}+\frac{1}{2}\|\mathbf{S}_{h}\mathbf{W}_{\psi_{h}}\mathbf{h}\|_{2}^{2}\right),
\end{equation}
which can also be sampled directly, given values for $\alpha_2$ and $\beta_2$.

\begin{algorithm}
    \caption{Gibbs sampling procedure}
    \label{alg:Gibbs_Scheme}
    \begin{algorithmic}[1]
        \REQUIRE The number of samples $n_{\text{samp}}$, initial guesses $\mathbf{g}^{0},\mathbf{h}^{0},\mathbf{\Lambda}^{0},\lambda_{h}^{0},$ and data $\mathbf{y}^{*}$.
        \FOR{$i=1,\ldots, n_{\text{samp}}$}
        \STATE //Sampling components
        \STATE $\mathbf{g}^{i}\sim \pi_{1}(\cdot|\mathbf{y}^{*},\mathbf{h}^{i-1},\mathbf{\Lambda}^{i-1})$: Sample by solving \eqref{eq:g_samp} using CGLS.
        \STATE $\mathbf{h}^{i}\sim \pi_{2}(\cdot|\mathbf{y}^{*},\mathbf{g}^{i},\lambda_{h}^{i-1})$: Sample by solving \eqref{eq:h_samp} using CGLS.
        \STATE // Sampling hyperparameters
        \STATE $\mathbf{\Lambda}^{i}\sim \pi_{3}(\cdot|\mathbf{g}^{i})$: Sample directly from an inverse gamma distribution defined in \eqref{eq:Hyperparameters_1}.
        \STATE $\lambda_{h}^{i}\sim \pi_{4}(\cdot|\mathbf{h}^{i})$: Sample directly from a gamma distribution defined in \eqref{eq:Hyperparameters_2}.
        \ENDFOR
        \RETURN $\{\mathbf{g}^{i},\mathbf{h}^{i},\mathbf{\Lambda}^{i},\lambda_{h}^{i}\}_{i=1}^{n_{\text{samp}}}$
    \end{algorithmic}
\end{algorithm}

\begin{remark}
    While $\mathbf{g}$ and $\mathbf{h}$ are structurally distinct as independent priors, they share a common dependency in the likelihood \eqref{eq:Likelihood}, making the posterior a correlated structure, in general. Without further intervention, this shared dependency allows the mean of the combined signal $\mathbf{f}$ to fluctuate between the two components during the sampling process. To resolve this identifiability issue, we frame the decomposition within a function space context by restricting $\mathbf{g}$ to a homogeneous space (specifically, mean-zero functions), while allowing $\mathbf{h}$ to accommodate the inhomogeneous, non-zero mean behavior. Computationally, at each iteration $i$ of the Gibbs sampler, we compute the re-centered samples as
    \begin{align*}
        \mathbf{\tilde{g}}^i= & \ \mathbf{g}^i-\bar{\mathbf{g}}^i, \\
        \mathbf{\tilde{h}}^i= & \ \mathbf{h}^i+\bar{\mathbf{g}}^i,
    \end{align*}
    where $\bar{\mathbf{g}}^i = \frac{1}{n} \sum_{j=1}^n (\mathbf{g}^i)_j$ denotes the spatial mean of the sample $\mathbf{g}^i$ obtained from Algorithm~\ref{alg:Gibbs_Scheme}. This re-centering ensures that $\mathbf{\tilde{g}}^i$ has zero-mean while the mean of $\mathbf{f}$ is consistently attributed to $\mathbf{\tilde{h}}^i$.
    
\end{remark}
\begin{remark} \label{Remark:2}
    The efficiency and behavior of the Gibbs sampler depend on the selection of the hyperprior-parameters $(\alpha_1, \beta_1)$ and $(\alpha_2, \beta_2)$. For the sparse component $\mathbf{g}$, the parameters $(\alpha_1, \beta_1)$ govern the Inverse Gamma distribution, which acts as a sparsity-promoting hyperprior on the gradients. To ensure that most gradients remain near zero while allowing for localized large transitions, we choose a small scale parameter $\beta_1=10^{-5}$ to shrink small fluctuations. Simultaneously, we set $\alpha_1 = 1$ to ensure a heavy-tailed distribution, which prevents the over-penalization of large gradient values. For the smooth component $\mathbf{h}$, the prior strength $\lambda_{h}$ is modeled with a Gamma hyperprior. Since we do not have prior information regarding the size of the smooth component and the noise level, we choose a weakly informative prior with $\alpha_2=2$ and $\beta_2 = 10^{-2}$ to allow the data to dictate the regularization strength.
\end{remark}

\section{Numerical experiments}
\label{sec:numer-exper}

In this section, we numerically evaluate the performance of the two proposed methods for reconstructing a signal composed of a piecewise constant component and a smooth component. 
We focus on linear deconvolution problems. The first two tests are one-dimensional problems, which allow for an in-depth investigation of the proposed methods. The third test is used to demonstrate the performance of the proposed hierarchical method on a two-dimensional problem.

\subsection{One dimensional deconvolution: Two-Besov prior method}
\label{sec:one-dimens-deconv}

We consider a one-dimensional periodic deconvolution problem, that is, we want to identify an unknown $f:\mathbb{T}\rightarrow \mathbb{R}$ from noisy convolved data. The periodic deconvolution problem can be modeled as an inverse problem
\begin{equation*}
    y(t)=\int_{\mathbb{T}}A(t-x)f(x) \,dx,
\end{equation*}
where $y$ is the observed data and $A\in \mathcal{C}^{\infty}(\mathbb{T})$ is a smooth convolution kernel. In this test, we use a $1$-periodic, centered Gaussian kernel with standard deviation $\sigma_{\text{ker}}>0$, i.e.,
\begin{equation}
    A(x)=\frac{1}{\sqrt{2\pi}\sigma_{\text{ker}}}\sum_{k=-\infty}^{\infty}\exp\left(\frac{-(x+k)^{2}}{2\sigma_{\text{ker}}^{2}}\right),\; x\in \mathbb{T}. \label{eq:GaussKernel}
\end{equation}

\begin{figure}[!htb]
    \centering
    \includegraphics[width=\linewidth]{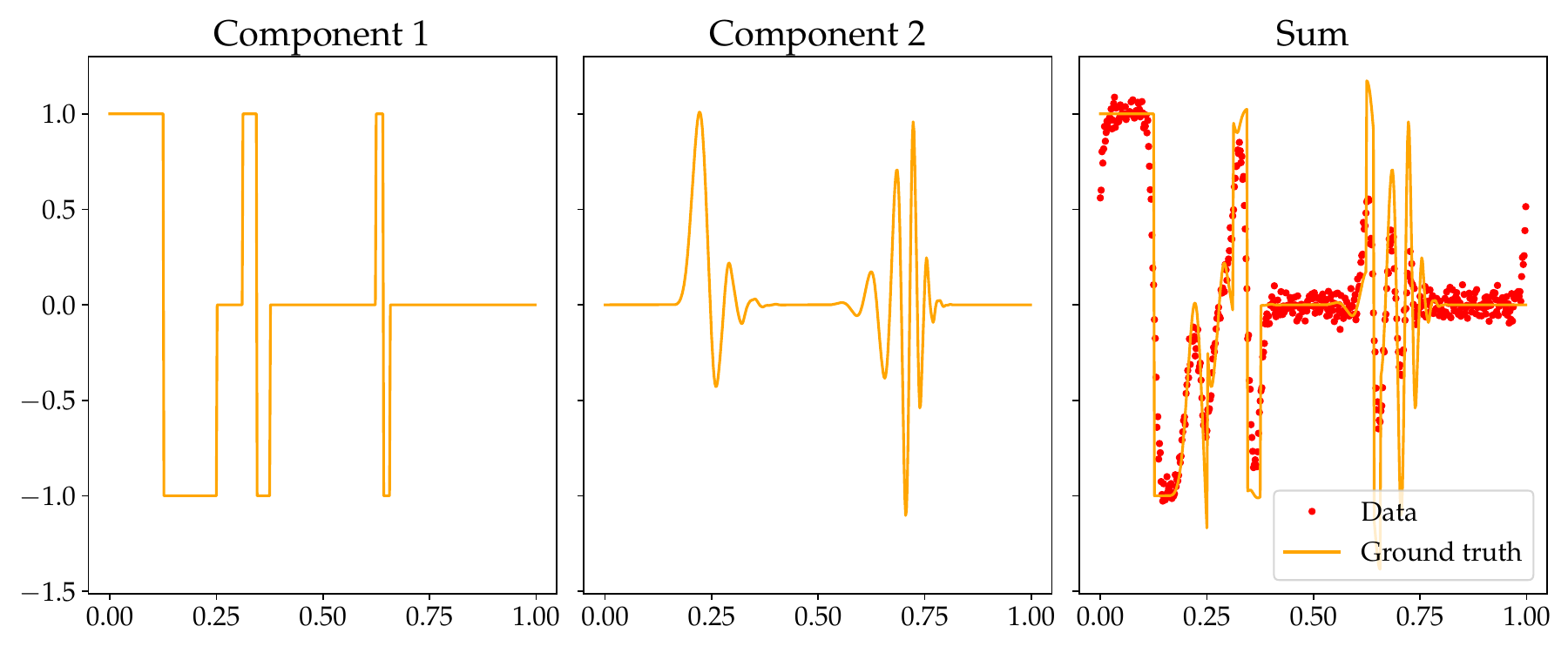}
    \caption{The ground truth signal and the degraded data (in red, to the right), together with the true piecewise constant component (to the left) and the smooth component (in the middle).}
    \label{fig:Test_1_Data}
\end{figure}

The ground truth signal $\mathbf{f}_{\text{true}}$ is constructed as a linear combination of Haar and DB(8) wavelets, discretized on a uniform grid of $n=512$ points over the interval $[0,1)$ with a truncation parameter $J=9$. We consider a spatially dense data regime ($m=n=512$) by discretizing the convolution operator $A$ as a circulant matrix $\mathbf{A} \in \mathbb{R}^{512 \times 512}$ on the same uniform grid as the signal. This matrix corresponds to the Gaussian convolution kernel in \eqref{eq:GaussKernel}, truncated at $\pm 3\sigma_{\text{ker}}$ with $\sigma_{\text{ker}}=0.01$. Noisy data $\mathbf{y}^*$ is generated by adding $10\%$ Gaussian noise $\epsilon$ to the blurred signal, such that $\|\mathbf{\epsilon}\|_{2}/\|\mathbf{A}\mathbf{f}_{\text{true}}\|_{2}=0.1$. The dense data setup with $n=m$ is maintained for all subsequent experiments. Fig.~\ref{fig:Test_1_Data} displays the ground truth components $\mathbf{g}$ and $\mathbf{h}$, the aggregate signal $\mathbf{f}_{\text{true}}$, and the resulting noisy observation $\mathbf{y}^*$.

In this test scenario, we study the performance of the method with two Besov priors introduced in Section~\ref{sec:BesovDecomposition}. We use the Haar wavelet $\psi_{g}$ for $\mathbf{g}$ and the DB(8) wavelet $\psi_{h}$ for $\mathbf{h}$. The prior parameters are set as $s_{g} = s_{h} = p_{g} = p_{h} = 1$. We first fix $\lambda_{g} = \lambda_{h} = 1$ to analyze the uncertainty quantification properties of the method and compare it to the method using a single Besov prior in terms of both sampling efficiency and reconstruction quality. While these fixed Besov priors are not tuned to achieve optimal reconstruction quality, they provide a controlled baseline that allows us to clearly assess the advantages and disadvantages of the decomposition model. We then study the influence of varying $\lambda_{g}$ and $\lambda_{h}$ on the reconstruction quality.

For all experiments in this test, inference is performed using NUTS to generate $\num{15000}$ total samples, of which the first $\num{5000}$ are discarded as burn-in. The remaining $\num{10000}$ samples are used to compute the posterior mean estimate, the 95\% credible interval (CI), and the CI width. To assess sampling quality, we compute the effective sample size (ESS) over the entire grid and report the minimum, median, and maximum values. Additionally, we evaluate the autocorrelation function (ACF) at a representative grid point to examine sample correlation. Reconstruction quality is measured using the relative error between the posterior mean and the ground truth $\mathbf{f}_{\text{true}}$.

\begin{figure}[!htb]
    \centering
    \includegraphics[width=\linewidth]{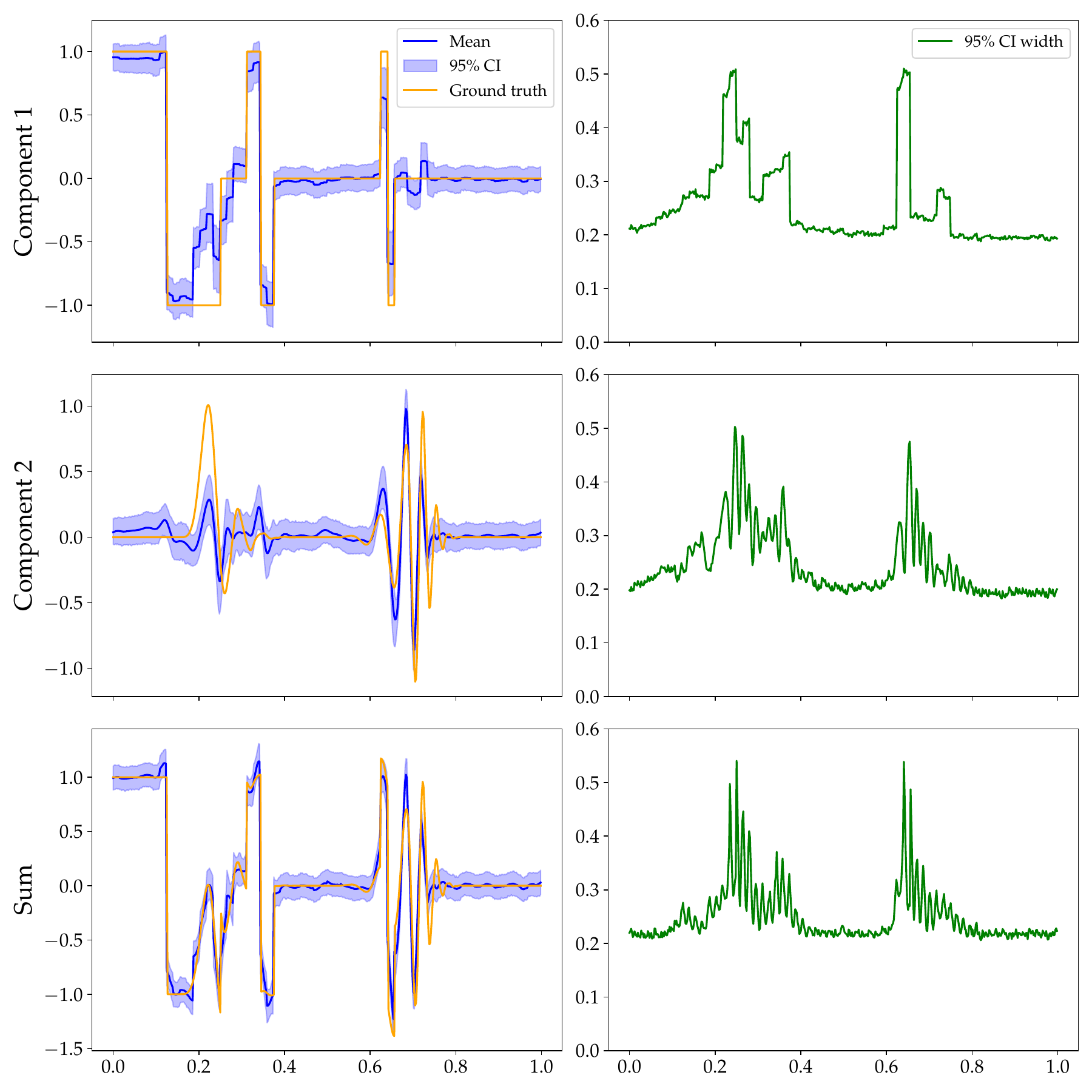}
    \caption{Posterior estimates using the Haar and DB(8) wavelets in the two-Besov prior decomposition model with the parameters $s_{g}=s_{h}=p_{g}=p_{h}=\lambda_{g}=\lambda_{h}=1$.}
    \label{fig:Test_1_Result_1}
\end{figure}

Fig.~\ref{fig:Test_1_Result_1} shows the reconstruction and uncertainty quantification results from the method with two Besov priors. The posterior mean estimates for $\mathbf{g}$ and $\mathbf{h}$ carry the most features from the true components, and the mean of $\mathbf{f}$, whose samples are obtained by calculating the sum of corresponding $\mathbf{g}$ and $\mathbf{h}$ samples, successfully represents both the sharp transitions and smooth variations present in the ground truth. The plots of the CI widths in Fig.~\ref{fig:Test_1_Result_1} show that uncertainty is higher in the regions where both components contribute significantly, indicating that the decomposition introduces additional variability in these areas.

\begin{figure}[!htb]
    \centering
    \includegraphics[width=\linewidth]{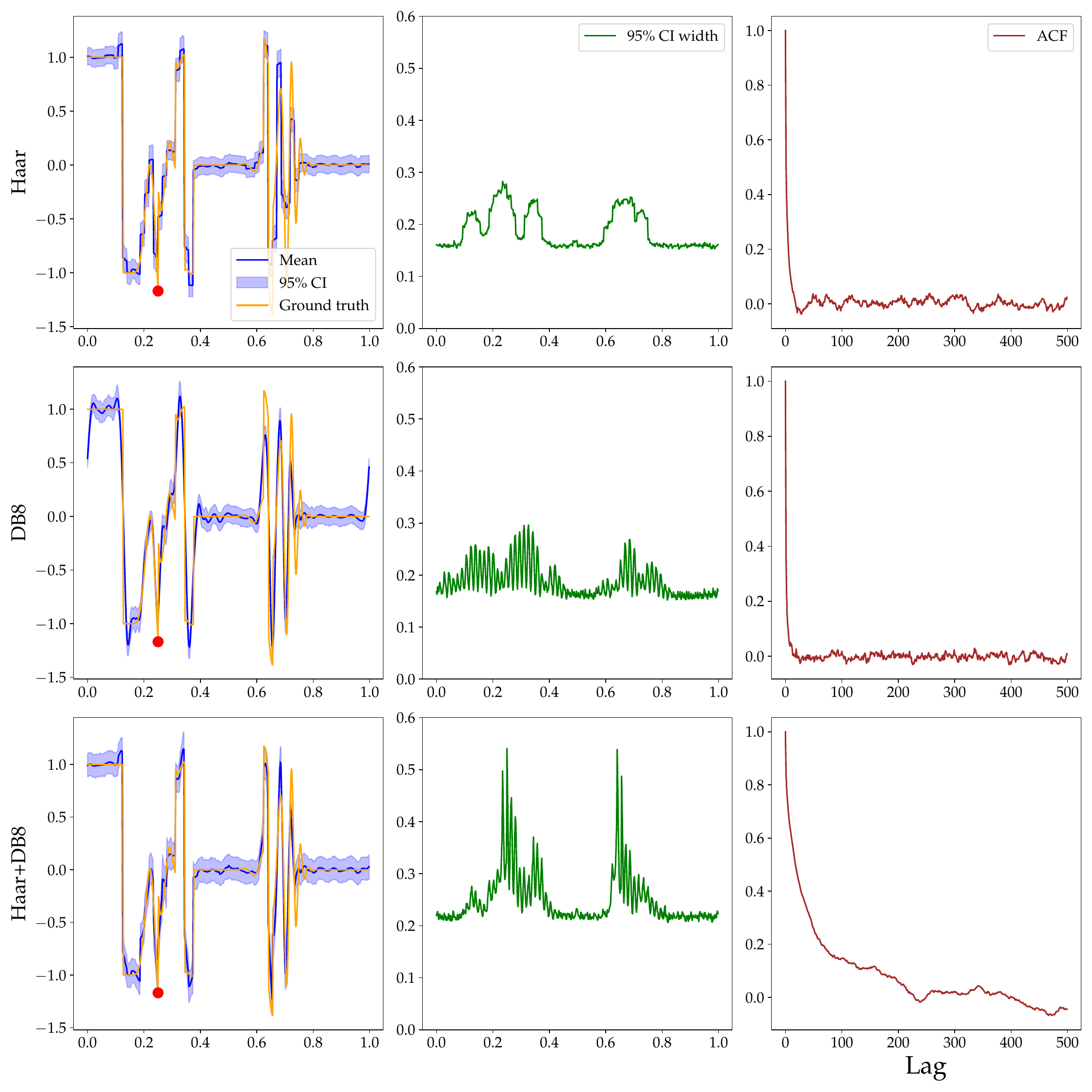}
    \caption{Posterior estimates and autocorrelation functions (ACFs) computed from the chains at a representative grid point (marked with a red dot). Row 1: One-Besov prior using Haar, i.e., DB(1), wavelets. Row 2: One-Besov prior using DB(8) wavelets. Row 3: Two-Besov prior model combining Haar and DB(8) wavelets (as in Section~\ref{sec:post-bayes-decomp}). All reconstructions use the same prior parameters $s=p=\lambda=1$ for single Besov priors and $s_{g}=s_{h}=p_{g}=p_{h}=\lambda_{g}=\lambda_{h}=1$ for the two-Besov model.}
    \label{fig:Test_1_Result_Compare}
\end{figure}

Next, we compare the performance of the two-Besov prior model (Haar+DB(8)) with that of standard single Besov priors (either Haar or DB(8) alone) applied directly to $\mathbf{f}$, using identical prior parameters and sampling settings as described above. As shown in Fig.~\ref{fig:Test_1_Result_Compare}, it is obvious that the Haar Besov prior introduces staircasing artifacts, while the DB(8) Besov prior tends to oversmooth the reconstruction. In contrast, the proposed two-Besov model achieves a more balanced reconstruction, mitigating both types of artifacts and yielding a lower relative reconstruction error, as summarized in Table~\ref{tab:Test_1}. Comparing the CI widths in Fig.~\ref{fig:Test_1_Result_Compare}, we can see that the results from the single Besov prior exhibit generally lower uncertainty compared to the two-Besov method, likely due to the inherent nonuniqueness and the mutual coherence between the two representation systems used in the decomposition. Moreover, the decomposition samples show higher correlation, as evidenced by slower ACF decay and lower ESS values. This indicates that sampling the posterior for decomposition problems is more challenging. Nevertheless, the correlation remains within acceptable limits, and the samples are deemed sufficient for a reliable posterior estimation.

\begin{table}[t]
    \caption{The effective sample sizes (ESS) and the relative errors.}
    \centering \label{tab:Test_1}
    \begin{tabular}{|c|ccc|c|}
        \hline
                                         & \multicolumn{3}{c|}{ESS}   & Relative                                                           \\ \cline{2-4}
        Method                           & Minimum                    & Median                      & Maximum                      & error \\ \hline
        Haar                             & 1318.68                    & 6906.55                     & 9047.97                      & 0.378 \\ \cline{1-1}
        DB(8)                            & 1296.79                    & 5947.35                     & 9135.38                      & 0.363 \\ \cline{1-1}
        \multicolumn{1}{|c|}{Haar+DB(8)} & \multicolumn{1}{c}{111.45} & \multicolumn{1}{c}{4500.12} & \multicolumn{1}{c|}{6077.02} & 0.308 \\ \hline
    \end{tabular}
\end{table}

\begin{figure}[!htb]
    \centering
    \includegraphics[width=\linewidth]{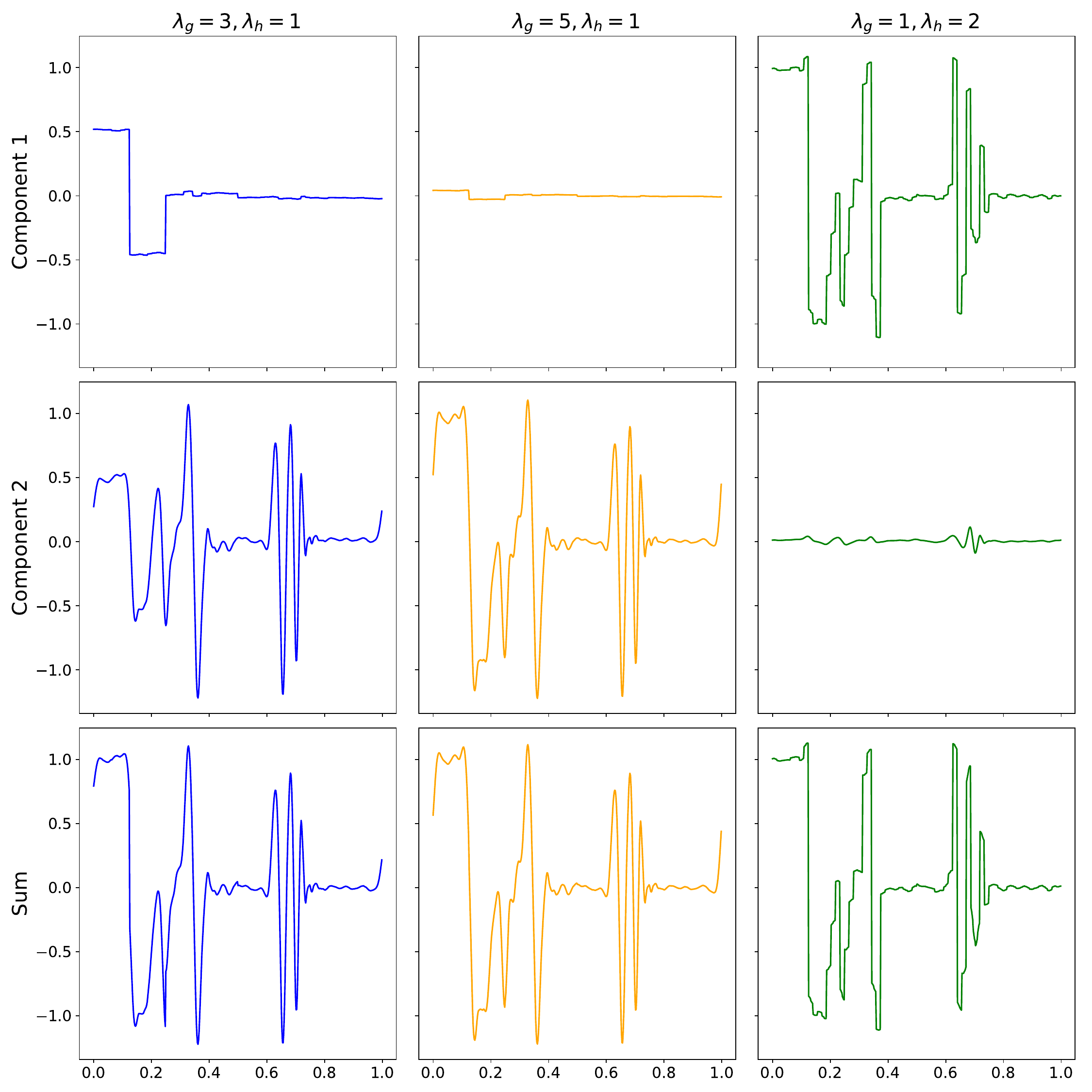}
    \caption{Posterior estimates for the decomposition method with different choices of prior strength parameters.}
    \label{fig:Test_1_Result_Parameter}
\end{figure}

In the two-Besov decomposition model, the parameters $\lambda_{g}$ and $\lambda_{h}$ control the relative contributions of the components in the posterior. Their choices have a substantial impact on the results as illustrated in Fig.~\ref{fig:Test_1_Result_Parameter}. As $\lambda_{g}$ increases, the second component progressively dominates the reconstruction, eventually becoming the primary contributor. Conversely, when $\lambda_{h}>\lambda_{g}$, the first component dominates. These results highlight the importance of selecting appropriate values for $\lambda_{g}$ and $\lambda_{h}$ to achieve a balanced and interpretable decomposition. This observation also motivates the second proposed method based on a hierarchical posterior.

\subsection{One dimensional deconvolution: Hierarchical decomposition method}
\label{sec:one-dimens-deconv-1}

\begin{figure}[!htb]
    \centering
    \includegraphics[width=\linewidth]{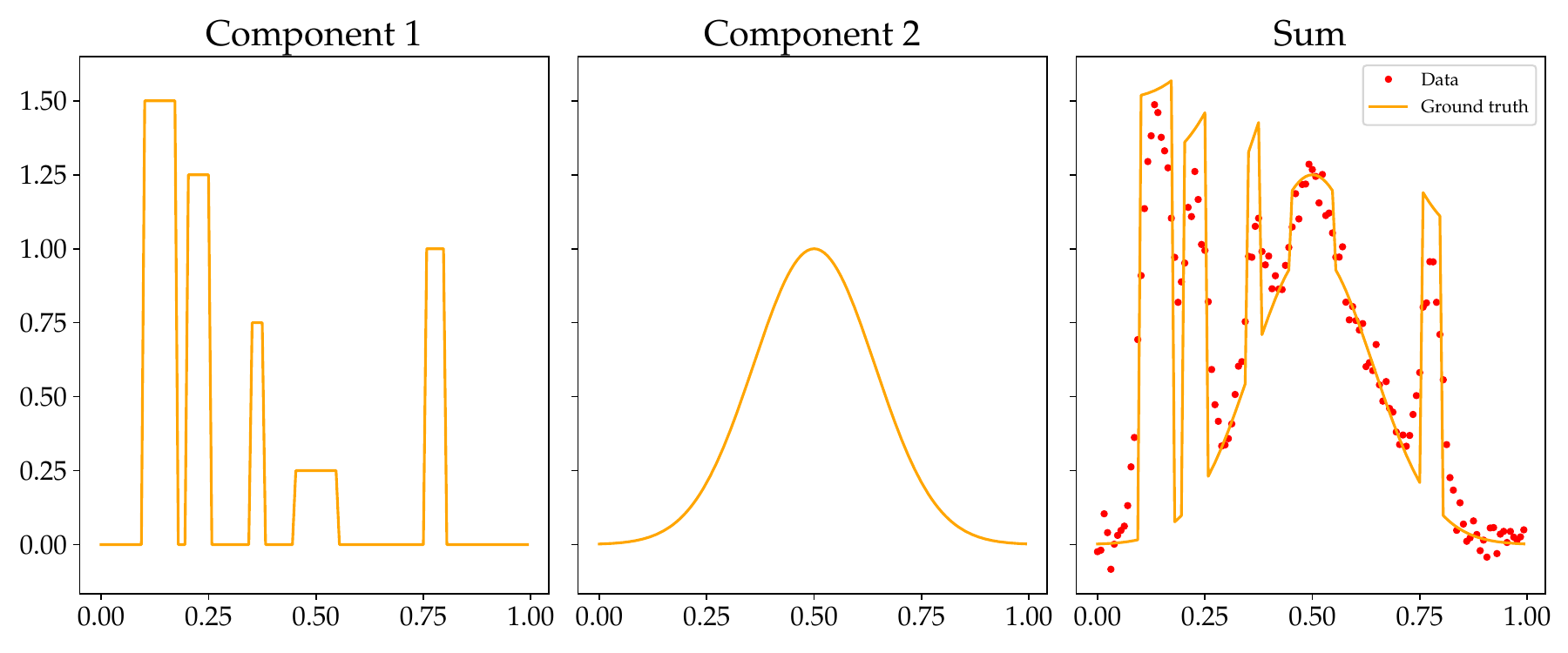}
    \caption{Ground truth (to the right) as a sum of the piecewise constant component (to the left) and the smooth component (in the middle) together with the degraded data (to the right).}
    \label{fig:Decomp_deconv_data_2}
\end{figure}

In this test scenario, the ground truth signal $\mathbf{f}_{\text{true}}$ is composed of a piecewise constant component and a smooth Gaussian component, both defined on the domain $[0,1)$ and discretized at $n=128$ equidistant points. The observed data is generated by convolving the signal with a periodic Gaussian kernel of standard deviation $\sigma_{\text{ker}}=0.02$, followed by the addition of $5\%$ Gaussian noise. This setup mirrors the dense data generation procedure described in Section~\ref{sec:one-dimens-deconv}.
A visualization of the ground truth and the observed data is shown in Fig.~\ref{fig:Decomp_deconv_data_2}.

First, we test the hierarchical decomposition method with two Besov priors on this problem. The hierarchical posterior and the associated sampling scheme are detailed in~\ref{sec:gibbs-sampl-hier}. We use the Haar wavelet for $\psi_{g}$ and the DB(8) wavelet for $\psi_{h}$, with prior smoothness parameters set to $s_{g} = 1$ and $s_{h} = 3$, respectively. In the Gibbs sampling scheme used for this experiment, the Besov prior strength parameters $\lambda_{g}$ and $\lambda_{h}$ are governed by Gamma hyperpriors with shape and rate parameters $(a_{1},b_{1})$ and $(a_{2}, b_{2})$, respectively. Specifically, we set $a_1 = a_2 = 2$ and $b_1 = b_2 = 10^{-3}$ to provide weakly informative hyperpriors. We draw $\num{1000000}$ samples, using the first $\num{10000}$ as burn-in and thinning the remainder by a factor of $\num{200}$, resulting in $\num{4950}$ samples for posterior analysis.

\begin{figure}[!htb]
    \centering
    \includegraphics[width=\linewidth]{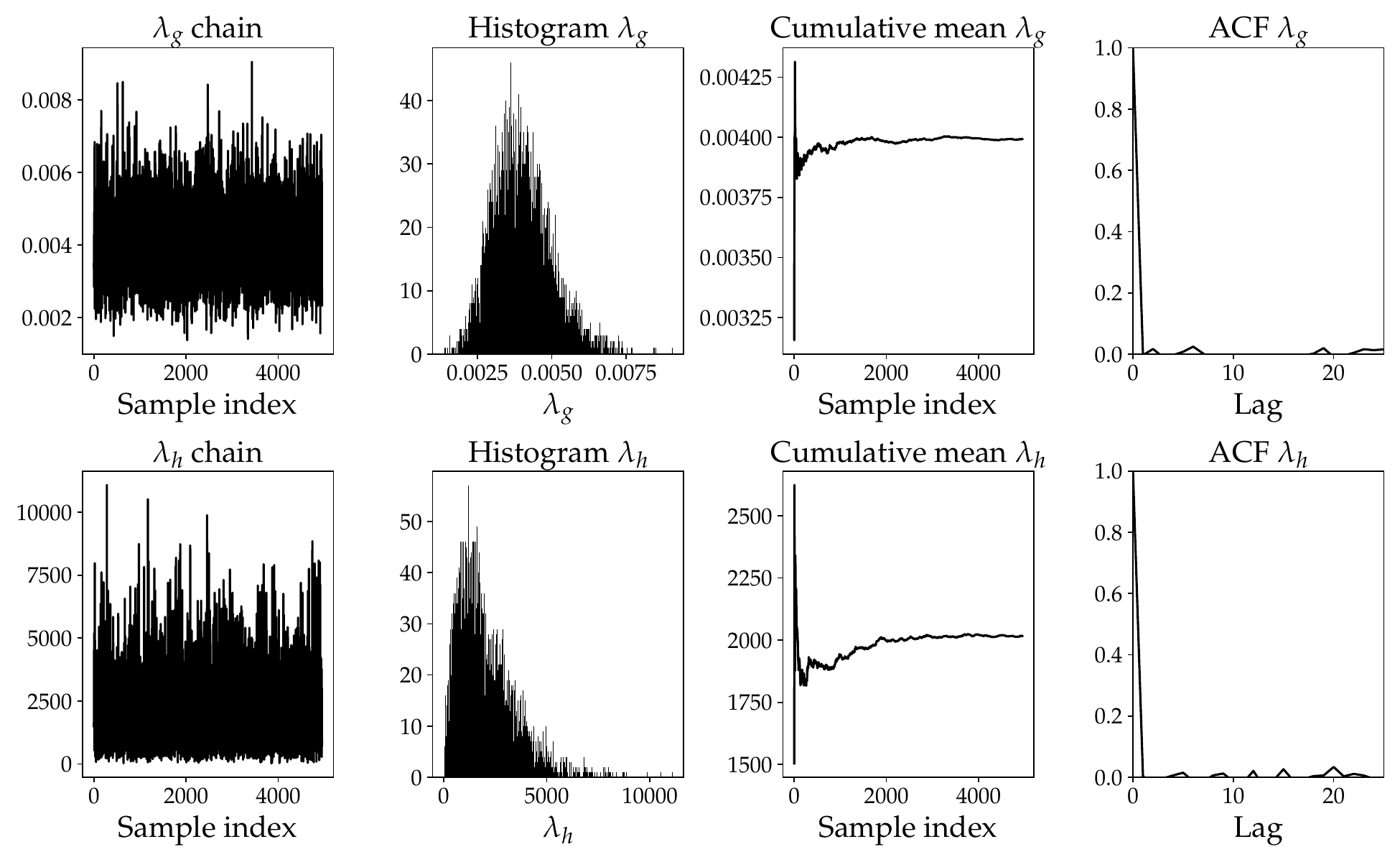}
    \caption{Posteriors of the hyperparameters $\lambda_{g}$ and $\lambda_{h}$ in the hierarchical decomposition method with two Besov priors.}
    \label{fig:Decomp_deconv_besov_param}
\end{figure}

\begin{figure}[thb!]
    \centering
    \includegraphics[width=\linewidth]{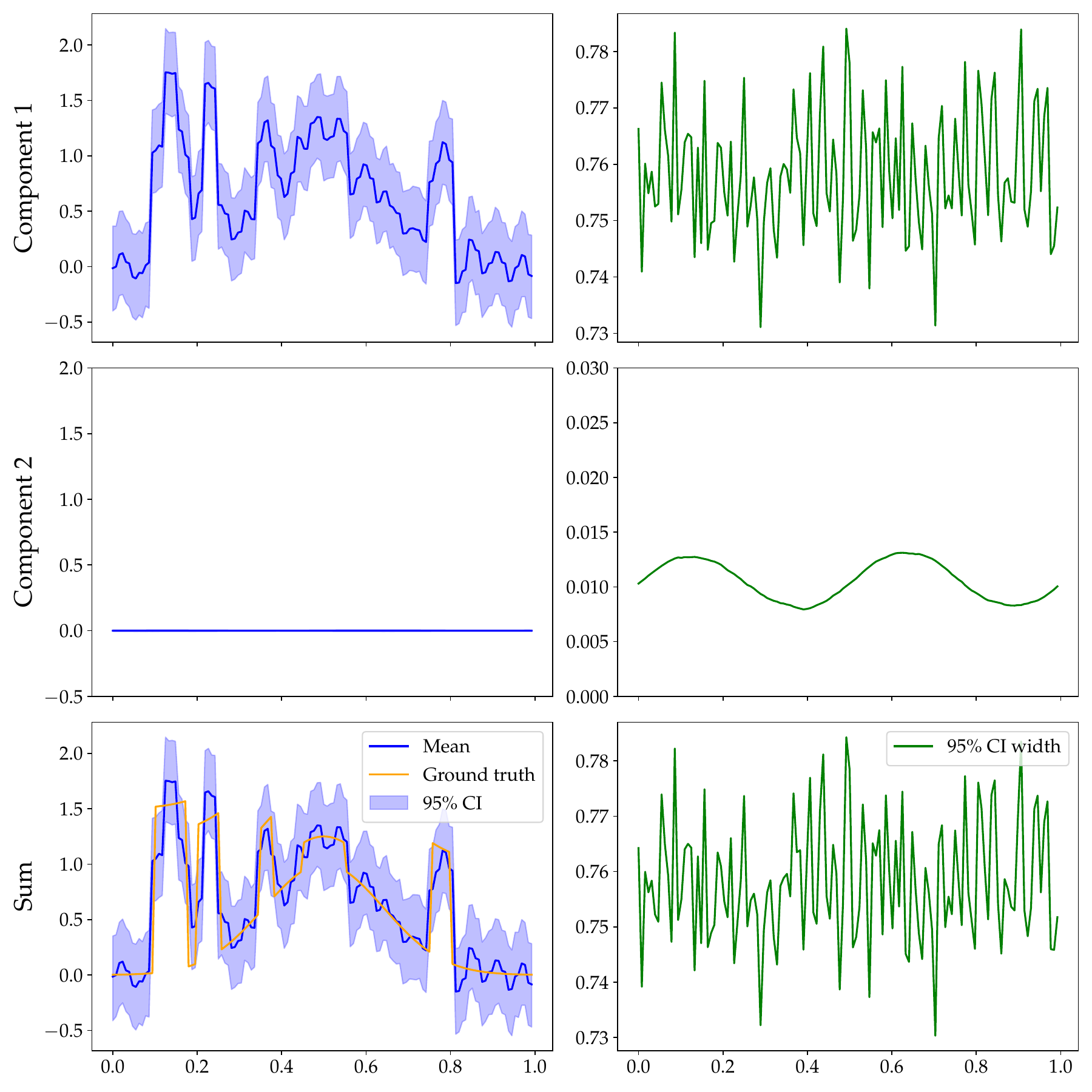}
    \caption{Posterior means and $95\%$ CI widths from the hierarchical decomposition method with two Besov priors.}
    \label{fig:Decomp_deconv_besov_x}
\end{figure}

In Fig.~\ref{fig:Decomp_deconv_besov_param} we show the diagnostics of $\lambda_{g}$ and $\lambda_{h}$ including individual chains, histograms, cumulative means and ACFs. We can see that both chains exhibit good mixing, cumulative means converge quickly, and ACFs reduce to zero rapidly. However, $\lambda_{g}$ is significantly smaller than $\lambda_{h}$, resulting in an imbalanced decomposition. This imbalance is evident in the posterior mean, where the component $\mathbf{g}$ dominates, as shown in Fig.~\ref{fig:Decomp_deconv_besov_x}. This test illustrates that a simply designed hierarchical decomposition method with two Besov priors fails due to identifiability limitations discussed in the end of Section~\ref{sec:BesovDecomposition}.

Next, we test the method proposed in Section~\ref{sec:hier-decomp-meth}, where we combine a hierarchical Gaussian prior with a Besov prior. For the Besov prior, we use DB(8) wavelet for $\psi_{h}$ and set $s_{h}=3$. The hyperprior-parameters $\alpha_1,\alpha_2$ and $\beta_1,\beta_2$ are chosen according to Remark~\ref{Remark:2}.
The number of samples, burn-in and thinning are the same as in the previous test.

\begin{figure}[!htb]
    \centering
    \includegraphics[width=\linewidth]{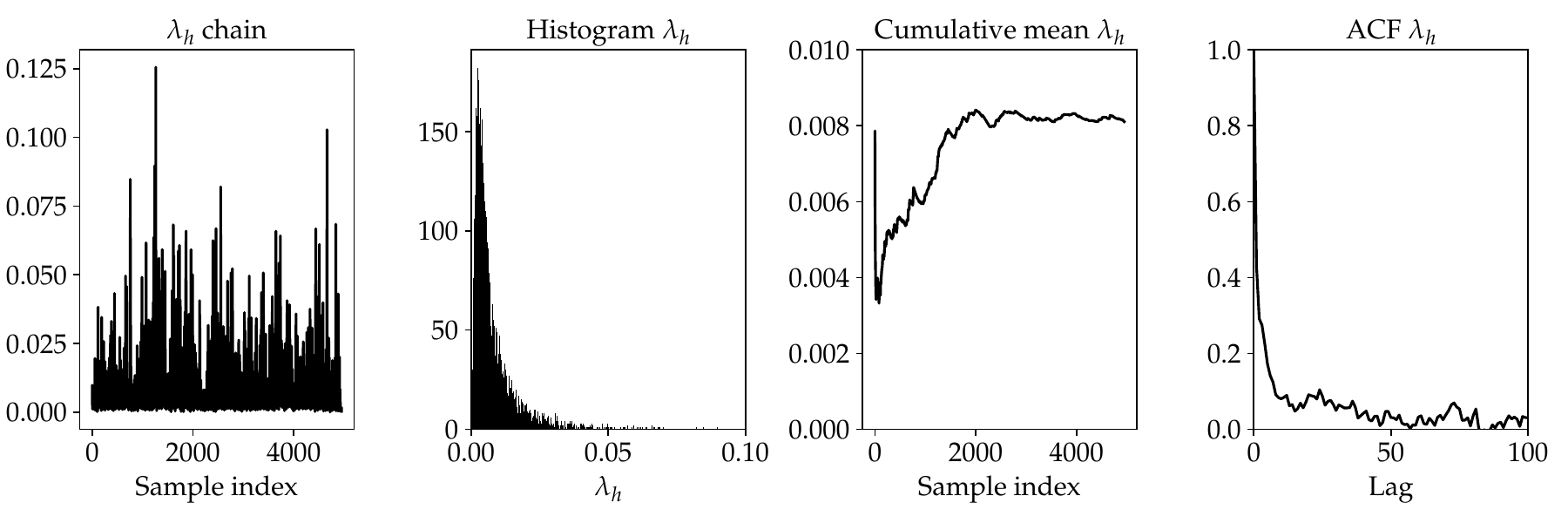}
    \caption{The posterior of the hyperparameter $\lambda_{h}$ in the hierarchical decomposition method proposed in Section~\ref{sec:hier-decomp-meth}.}
    \label{fig:Decomp_deconv_hierarchical_param}
\end{figure}

\begin{figure}[!htb]
    \centering
    \includegraphics[width=\linewidth]{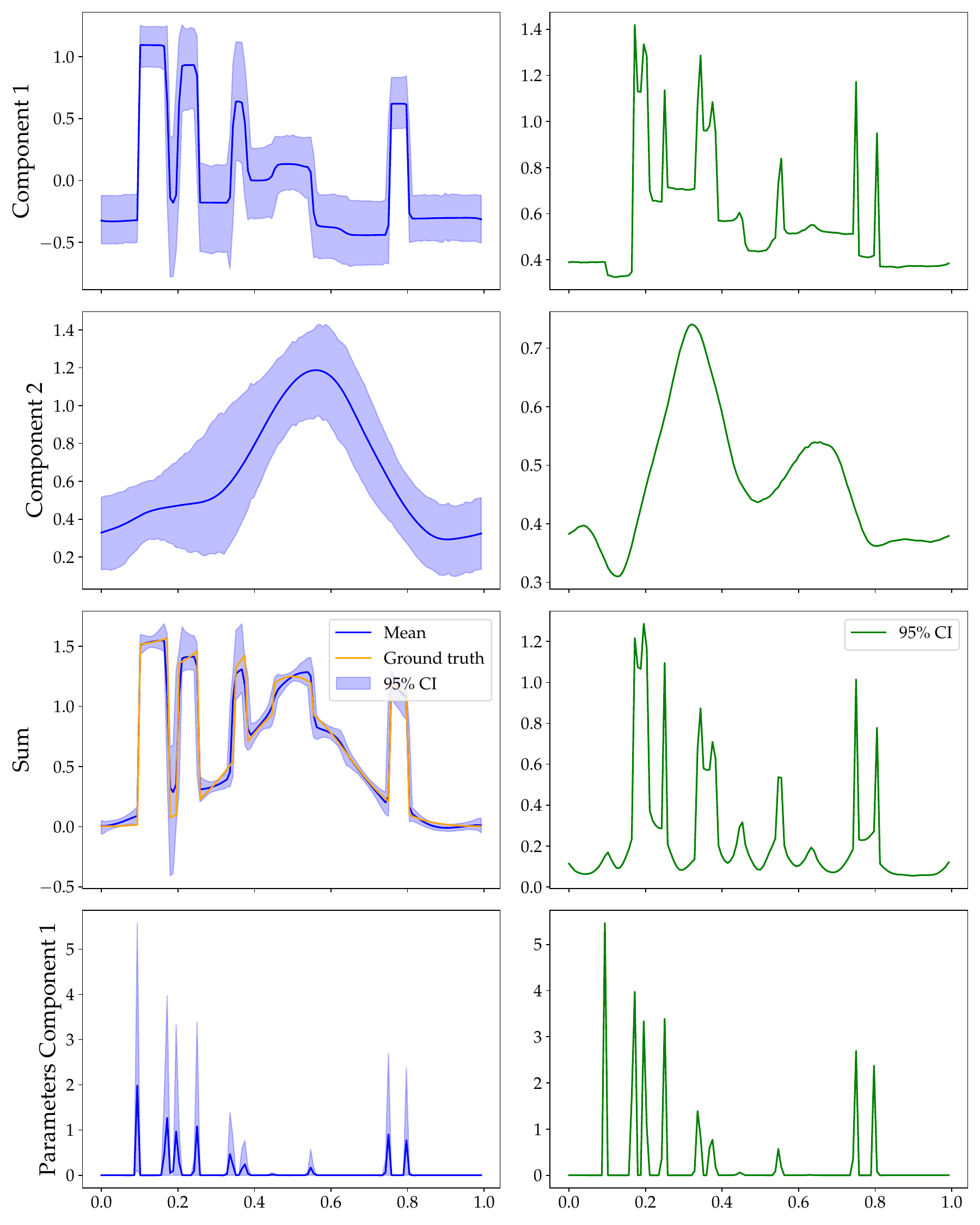}
    \caption{Posterior means and $95\%$ CI widths in the hierarchical decomposition method introduced in Section~\ref{sec:hier-decomp-meth}.}
    \label{fig:Decomp_deconv_hierarchical_x}
\end{figure}

In Fig.~\ref{fig:Decomp_deconv_hierarchical_param} we show the posterior statistics of $\lambda_{h}$. Comparing with the posterior statistics of hyperparameters in two Besov prior case which is shown in Fig.~\ref{fig:Decomp_deconv_besov_param}, we can see that the chain exhibits some correlation with occasional outliers, and the cumulative mean converges slower. The posterior estimates of both components and their sum $\mathbf{f}$ are shown in Fig.~\ref{fig:Decomp_deconv_hierarchical_x}. We observe that the component $\mathbf{g}$ captures nearly all ten jumps, estimating both their locations and magnitudes with reasonable accuracy. Furthermore, the posterior mean of $\mathbf{f}$ provides a qualitatively accurate reconstruction of the unknown. According to the $95\%$ CI width of $\mathbf{f}$, we see that higher uncertainties are observed near the jumps, reflecting the difficulty of reconstructing high-frequency information. In the last row of Fig.~\ref{fig:Decomp_deconv_hierarchical_x}, we also show the posterior statistics for $\mathbf{\Lambda}$, the covariance of the hierarchical Gaussian prior. It is obvious that $\mathbf{\Lambda}$ is sparse, and the locations of non-zero elements correspond to the jumps in the gradient. The uncertainty is large at those particular locations and essentially zero in the rest of the domain.  

Finally, we quantitatively compare two hierarchical decomposition methods introduced in~\ref{sec:gibbs-sampl-hier} with two Besov priors and in Section~\ref{sec:hier-decomp-meth} with a hierarchical Gaussian and Besov prior. The results are summarized in Table~\ref{Tab:test_2}. While the method with two Besov priors provides superior sampling efficiency, as measured by ESS of the hyperparameters, it fails to produce a balanced decomposition.
In contrast, the method with a hierarchical Gaussian and a Besov prior is less efficient on sampling, but it successfully identifies hyperparameters that yield a decomposition with substantially lower relative error.

\begin{table}[t]
    \caption{Comparison of two hierarchical decomposition methods.}
    \centering
    \begin{tabular}{|ccc|cll|}
        \hline
        \multicolumn{3}{|c|}{Two Besov priors}  & \multicolumn{3}{c|}{Hierarchical Gaussian with Besov}                                                                                                                    \\ \hline
        \multicolumn{1}{|c|}{ESS $\lambda_{g}$} & \multicolumn{1}{c|}{ESS $\lambda_{h}$}                & Rel\_err & \multicolumn{1}{c|}{Min ESS $\mathbf{\Lambda}$} & \multicolumn{1}{l|}{ESS $\lambda_{h}$} & Rel\_error \\ \hline
        \multicolumn{1}{|c|}{4893.45}           & \multicolumn{1}{c|}{4883.70}                          & 0.307    & \multicolumn{1}{c|}{22.80}                      & \multicolumn{1}{l|}{335.61}            & 0.103      \\ \hline
    \end{tabular}
    \label{Tab:test_2}
\end{table}

\subsection{Image deblurring}
\label{sec:image-deblurring}
\begin{figure}[!htb]
    \centering
    \includegraphics[width=\linewidth]{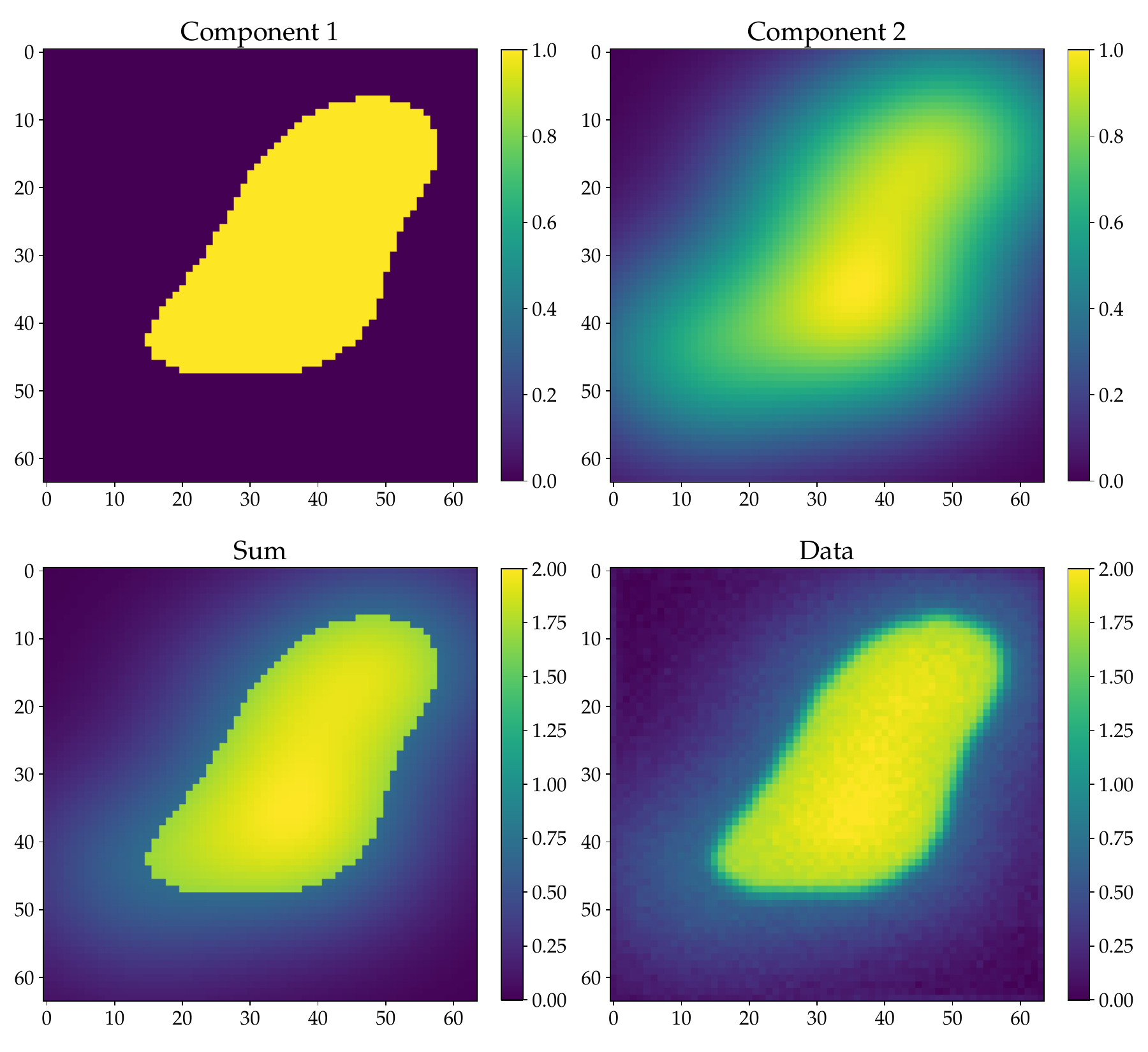}
    \caption{Ground truth and degraded image for the 2D deconvolution problem.}
    \label{fig:decomp_deblur_data}
\end{figure}

In this section, we apply our hierarchical decomposition method proposed in Section~\ref{sec:hier-decomp-meth} on a 2D deconvolution problem. We consider a separable Gaussian convolution kernel, i.e., the convolution is performed through a 1D Gaussian convolution defined in Section~\ref{sec:one-dimens-deconv} along both the horizontal and vertical directions. The ground truth $\mathbf{f}_{\text{true}}$ comprises a piecewise constant component and a smooth component, discretized on a $64 \times 64$ grid for a total of $n = 4096$ pixels with intensities in the range $[0, 2]$. To generate the observed data $\mathbf{y}^*$, we apply the blur kernel with $\sigma_{\text{ker}}=1.0$ and add Gaussian noise at a $2\%$ relative noise level. This results in a spatially dense dataset where the number of observations $m$ matches the signal dimension ($m=n=4096$). The ground truth image and its degraded counterpart are illustrated in Fig.~\ref{fig:decomp_deblur_data}.

In our method, we use the DB(8) wavelet with $s_{h}=3$ in the Besov prior for $\psi_{h}$. The hyperprior-parameters $\alpha_1,\alpha_2$ and $\beta_1,\beta_2$ are chosen based on Remark~\ref{Remark:2}.

We run Algorithm~\ref{alg:Gibbs_Scheme} to generate $\num{200000}$ samples, from which we discard the first $\num{10000}$ as burn-in and thin the remaining samples by a factor of 30, providing $\num{6334}$ samples used for analysis.

\begin{figure}[!htb]
    \centering
    \includegraphics[width=\linewidth]{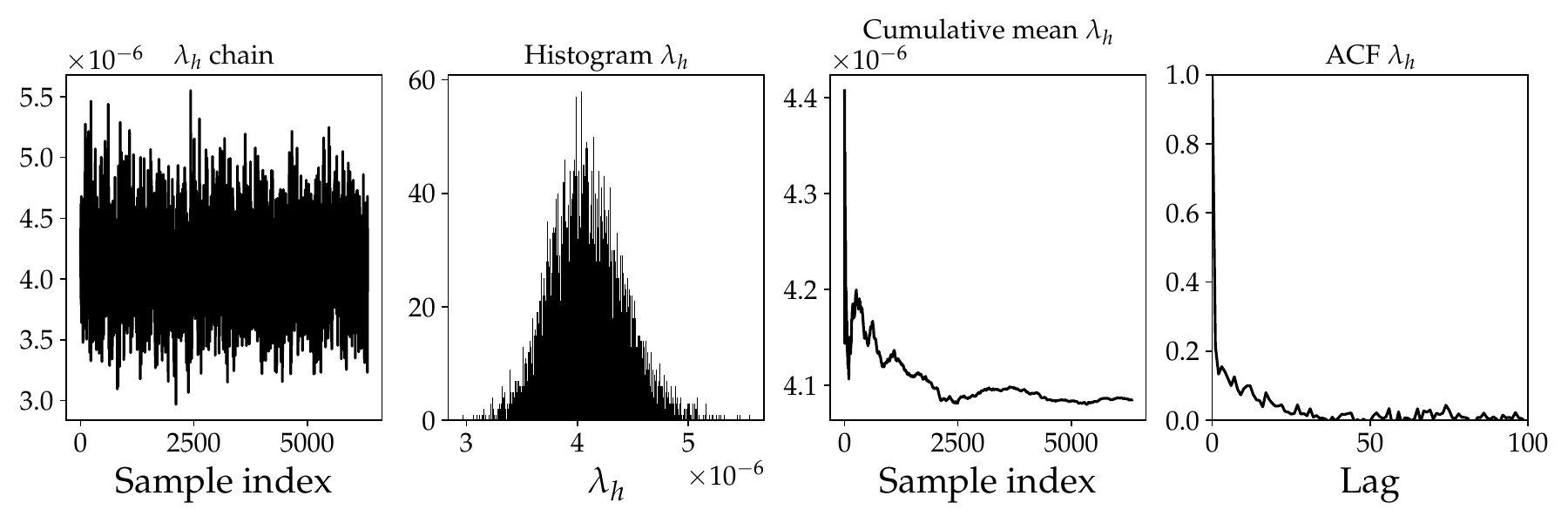}
    \caption{Posterior statistics of the hyperparameter $\lambda_{h}$ from the hierarchical decomposition method for the 2D deconvolution problem.}
    \label{fig:Decomp_deblur_param_1}
\end{figure}

\begin{figure}[!htb]
    \centering
    \includegraphics[width=\linewidth]{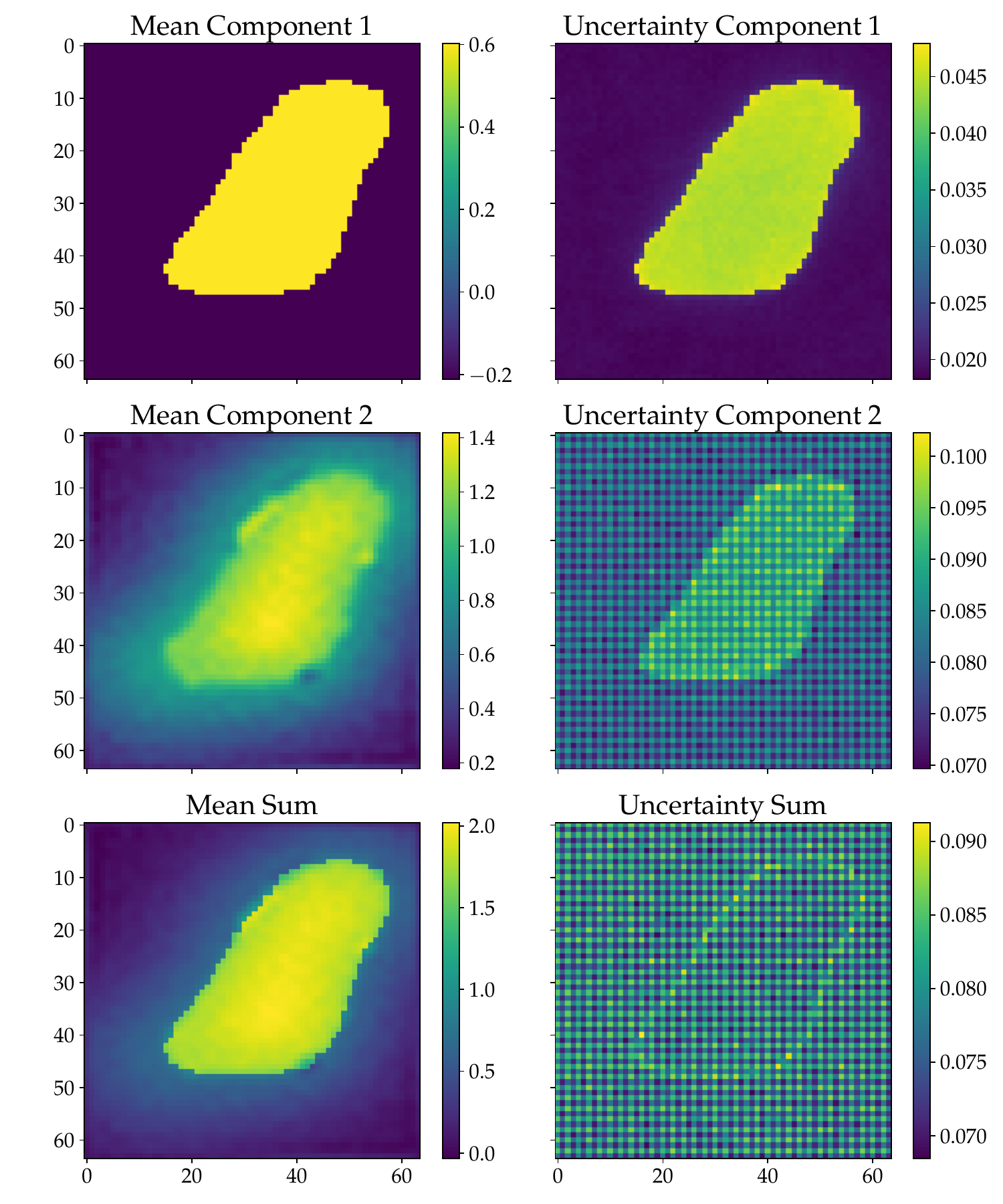}
    \caption{Posterior means and the pixel-wise $95\%$ CI widths of $\mathbf{g}$, $\mathbf{h}$ and $\mathbf{f}$ for the 2D deconvolution problem.}
    \label{fig:Decomp_deblur_X}
\end{figure}

Fig.~\ref{fig:Decomp_deblur_param_1} shows the posterior statistics for the hyperparameter $\lambda_{h}$. The chain appears well-mixed, the cumulative mean converges quickly, and the autocorrelation is low, indicating that the samples are reliable for posterior inference.
In Fig.~\ref{fig:Decomp_deblur_X} we show the posterior mean and the pixel-wise $95\%$ CI width. The posterior mean of $\mathbf{g}$ accurately captures the piecewise constant structure, and the posterior mean of $\mathbf{h}$ reflects the smooth variations. According to the colorbar, we can see that a part of the piecewise constant component was decomposed into $\mathbf{h}$, which indicates that an even smoother prior would be preferred for $\mathbf{h}$. Nevertheless, the posterior mean of $\mathbf{f}$ is close to the ground truth with correct intensity range and the relative error of $\num{0.06}$.

In the uncertainty plots of both $\mathbf{h}$ and $\mathbf{f}$ in Fig.~\ref{fig:Decomp_deblur_X}, we observe a regular, periodic pattern induced by the Besov prior. This structure is a natural consequence of the prior's construction: when the Besov smoothness parameter $s_h \neq 0$, the scaling matrix $\mathbf{S}_h$ is no longer a scalar multiple of the identity. Instead, it introduces scale-dependent weights, which result in anisotropic uncertainty. Such patterns are inherent to any non-trivial Besov prior and may appear more clearly in two-dimensional problems. In one dimension, or in problems where the likelihood dominates the prior, the effect may go unnoticed. A more detailed explanation is provided in~\ref{sec:besov-patterns}.
 
However, ignoring these structural patterns, we still see that the higher uncertainties
are mainly at the boundary of the object in $\mathbf{f}$.

\begin{figure}[!htb]
    \centering
    \includegraphics[width=\linewidth]{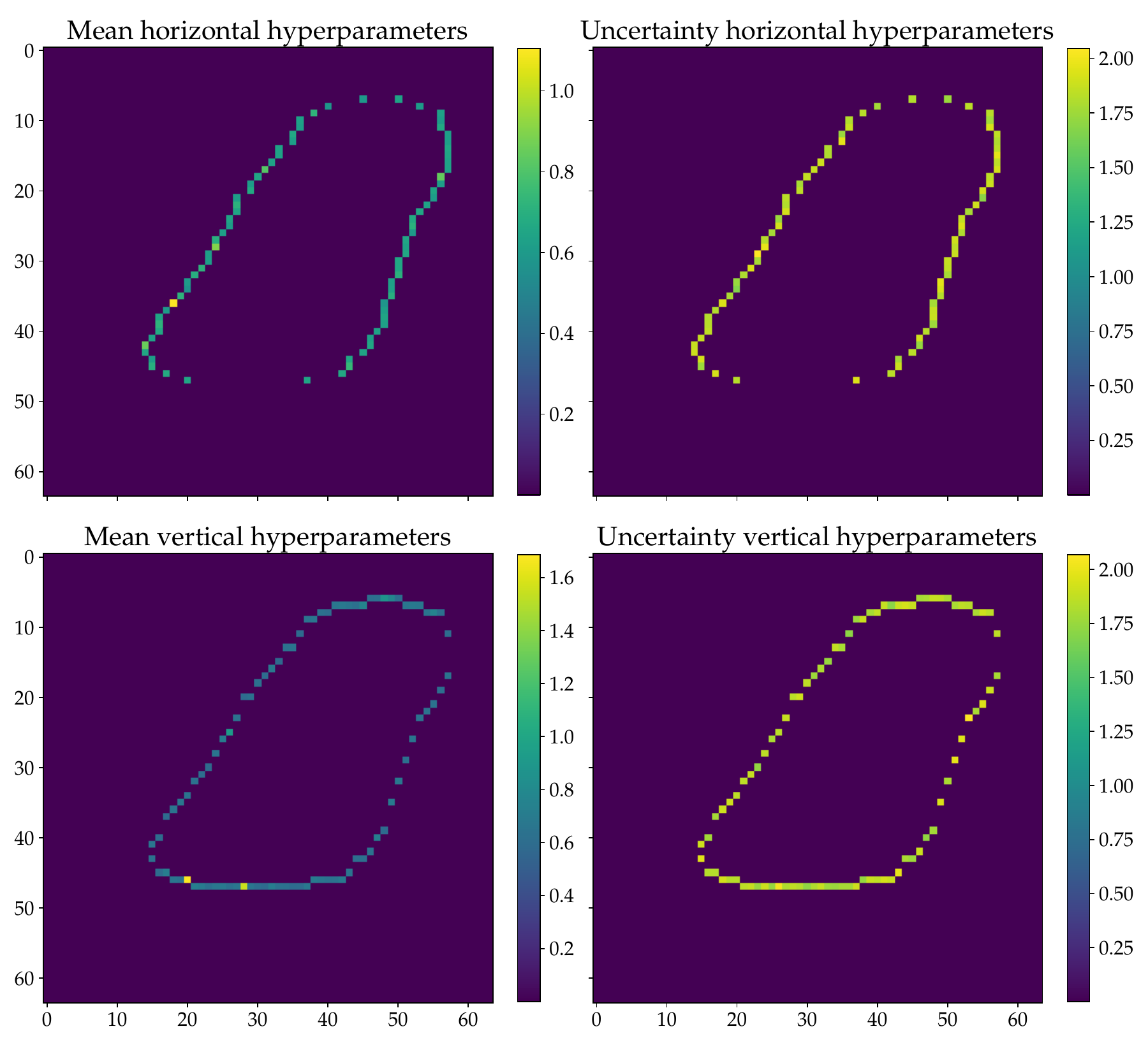}
    \caption{Posterior statistics of the hyperparameter $\mathbf{\Lambda}$ for the 2D deconvolution problem. Row 1: pixel-wise mean and STDs associated to horizontal derivative. Row 2: pixel-wise mean and STDs associated to vertical derivative. }
    \label{fig:Decomp_deblur_param_2}
\end{figure}

Fig.~\ref{fig:Decomp_deblur_param_2} shows the posterior means and the pixel-wise standard deviations (STDs) of the main diagonal in $\mathbf{\Lambda}$, which correspond to STDs of the first order derivatives of $\mathbf{g}$.
It is clear that the posterior means are significant only near the boundaries, resulting in a sparse structure. These boundary regions also correspond to areas with noticeable high uncertainties.

\section{Conclusion}
\label{sec:conclusion}
In this paper, we have developed computational methods for solving linear Bayesian inverse problems where the unknown consists of smooth regions connected by localized jumps. Our approach is based on modeling the unknown as a sum of two components that individually describes either the smooth part or the jumps represented by a piecewise constant component. We propose two prior models that assign different priors to each component: the first combines two Besov priors, while the second uses a hierarchical Gaussian prior for the piecewise constant component and a smooth Besov prior for the smooth part.

We also address the challenge of selecting suitable hyperparameters for each prior to ensure a balanced decomposition where both components contribute as desired. We use a hierarchical Bayesian approach and put hyperpriors on the parameters, and we propose a Gibbs sampling framework to infer the resulting posterior distribution.

From numerical experiments on 1D and 2D deconvolution problems, we conclude the following:
\begin{itemize}
    \item The two-Besov prior model yields better reconstructions compared to a single Besov prior model, but sampling using NUTS results in more correlated posterior samples.
    \item In the two-Besov model, the balance between components is highly sensitive to the prior strength parameters.
    \item The hierarchical model with two Besov priors estimates the components and the prior strength parameters simultaneously. However, because of the identification issue, it often results in one component dominating the other, leading to unbalanced reconstructions.
    \item The hierarchical Gaussian-Besov decomposition model finds a parameter setting that achieves a balanced solution but with high correlation between the samples.
\end{itemize}
While our experiments focus on spatially dense data ($m=n$), the proposed framework is inherently suited for sparse data regimes where $m < n$. In such sparse settings, the likelihood is less informative, and the reconstruction relies more on the priors to act as regularizers. We believe that our hierarchical model is particularly advantageous here, as it automatically adapts the prior parameters to compensate for missing observations, effectively using the assumed regularity of $\mathbf{g}$ and $\mathbf{h}$ to resolve the components in unobserved regions.

Although the hierarchical Gaussian prior does not strictly satisfy the notion of discretization-invariance as defined in \cite{Lassas2009}, it still provides highly sparse gradients, which are advantageous in practice. Consequently, the accuracy and efficiency of the proposed method may exhibit some dependence on the discretization resolution. Likewise, the sampling strategies used in this work, NUTS and Gibbs, are primarily designed for finite-dimensional settings, and their performance can be influenced by mesh refinement \cite{Cotter2013MCMC}. Nevertheless, these aspects do not hinder the effectiveness of our approach in the presented examples, where the discretization is fixed and of moderate size. Extending this framework toward discretization-invariant decomposition models and incorporating infinite-dimensional Markov chain Monte Carlo methods represents an interesting avenue for future research.

Extending the framework presented in this work to nonlinear forward operators presents a compelling direction for future research. The main findings established here, including the Gibbs sampling scheme in Algorithm~\ref{alg:Gibbs_Scheme}, can be naturally extended to a nonlinear setting using a nonlinear RTO approach \cite{Bardsley2014}. However, because evaluating the determinant of the linearized mapping within the nonlinear RTO framework is often computationally intensive, developing efficient computational strategies for this step is an interesting research direction.

\section*{Acknowledgments}
This work was supported by the Villum Investigator Grant (no.\ 25893) from the Villum Foundation. Babak Maboudi Afkham is partially supported by the Research Council of Finland project numbers: 359186, 353093.

\appendix

\section{Gibbs sampler for the hierarchical decomposition model with two Besov priors}
\label{sec:gibbs-sampl-hier}

Here, we consider a hierarchical formulation in which the prior strength parameters $\lambda_g$ and $\lambda_h$ in \eqref{eq:Posterior_1} are treated as random variables with associated hyperpriors. This approach follows the same Bayesian framework as the model in Section~\ref{sec:hier-decomp-meth}. However, it suffers from identifiability issues which can lead to imbalanced decompositions, where one component dominates. We include the sampling scheme here for completeness and comparison.

We assume that both $\lambda_{g}$ and $\lambda_{h}$ follow Gamma distributions defined as
\begin{equation*}
    \pi_{\text{hyp}}(\lambda_{g})\propto \lambda_{g}^{a_{1}-1}\exp\left(-b_{1}\lambda_{g}\right), \quad \pi_{\text{hyp}}(\lambda_{h})\propto \lambda_{h}^{a_{2}-1}\exp\left(-b_{2}\lambda_{h}\right),
\end{equation*}
where $a_1$, $b_1$, $a_2$ and $b_2$ are all positive.
Then the joint posterior of the components and hyperparameters is
\begin{align}
    \pi_{\text{post}}(\mathbf{g},\mathbf{h},\lambda_{g},\lambda_{h}|\mathbf{y}=\mathbf{y}^{*})
     & \propto \pi_{\text{like}}(\mathbf{y}=\mathbf{y}^{*}|\mathbf{g},\mathbf{h})\pi_{\text{prior}}(\mathbf{g}|\lambda_{g})\pi_{\text{prior}}(\mathbf{h}|\lambda_{h})\pi_{\text{hyp}}(\lambda_{g})\pi_{\text{hyp}}(\lambda_{h}) \nonumber              \\
     & \propto \lambda_{g}^{n/p_{g}+a_{1}-1} \lambda_{h}^{n/p_{h}+a_{2}-1} \exp\Bigl(-\frac{1}{2\sigma^{2}} \|\mathbf{A}(\mathbf{g}+\mathbf{h})-\mathbf{y}^{*}\|_{2}^{2} \nonumber                                                                     \\
     &-\lambda_{g} \|\mathbf{S}_{g}\mathbf{W}_{\psi_{g}}\mathbf{g}\|_{p_{g}}^{p_{g}} - \lambda_{h}\|\mathbf{S}_{h}\mathbf{W}_{\psi_{h}}\mathbf{h}\|_{p_{h}}^{p_{h}} - b_{1}\lambda_{g} - b_{2}\lambda_{h}\Bigr). \label{eq:PosteriorBesovHierarhical}
\end{align}
The posterior in \eqref{eq:PosteriorBesovHierarhical} can be sampled using a Gibbs sampling method similar to Algorithm~\ref{alg:Gibbs_Scheme} on four marginal distributions:
\begin{subequations} \label{eq:BesovMarginals}
    \begin{align*}
        \pi_{1}(\mathbf{g}|\mathbf{y}^{*},\lambda_{g},\mathbf{h}) & \propto \exp\left(-\frac{1}{2\sigma^{2}}\|\mathbf{A}(\mathbf{g}+\mathbf{h})-\mathbf{y}^{*}\|_{2}^{2}-\lambda_{g}\|\mathbf{S}_{g}\mathbf{W}_{\psi_{g}}\mathbf{g}\|_{p_{g}}^{p_{g}}\right) \\
        \pi_{2}(\mathbf{h}|\mathbf{y}^{*},\lambda_{h},\mathbf{g}) & \propto \exp\left(-\frac{1}{2\sigma^{2}}\|\mathbf{A}(\mathbf{g}+\mathbf{h})-\mathbf{y}^{*}\|_{2}^{2}-\lambda_{h}\|\mathbf{S}_{h}\mathbf{W}_{\psi_{h}}\mathbf{h}\|_{p_{h}}^{p_{h}}\right) \\
        \pi_{3}(\lambda_{g}|\mathbf{g})                           & \propto \lambda_{g}^{n/p_{g}+a_{1}-1}\exp\left(-\lambda_{g}\|\mathbf{S}_{g}\mathbf{W}_{\psi_{g}}\mathbf{g}\|_{p_{g}}^{p_{g}}-b_{1}\lambda_{g}\right)                                     \\
        \pi_{4}(\lambda_{h}|\mathbf{h})                           & \propto \lambda_{h}^{n/p_{h}+a_{2}-1}\exp\left(-\lambda_{h}\|\mathbf{S}_{h}\mathbf{W}_{\psi_{h}}\mathbf{h}\|_{p_{h}}^{p_{h}}-b_{2}\lambda_{h}\right)
    \end{align*}
\end{subequations}
We set $p_{g}=p_{h}=2$ and use RTO to sample $\pi_{1}$ and $\pi_{2}$. The marginal distributions $\pi_{3}$ and $\pi_{4}$ can be sampled directly according to Gamma distributions defined as follows
\begin{align*}
        \lambda_{g}^{\text{sample}} & \sim \mathcal{G}(a_{1}+n/2,b_{1}+\|\mathbf{S}_{g}\mathbf{W}_{\psi_{g}}\mathbf{g}\|_{2}^{2}),\\
        \lambda_{h}^{\text{sample}} & \sim \mathcal{G}(a_{2}+n/2,b_{2}+\|\mathbf{S}_{h}\mathbf{W}_{\psi_{h}}\mathbf{h}\|_{2}^{2}).
\end{align*}

\section{Anisotropic uncertainty in Besov priors}
\label{sec:besov-patterns}

We consider the case where the wavelet coefficients $\mathbf{c} \in \mathbb{R}^n$ are
drawn from a product distribution of identically distributed generalized
Gaussian random variables with density \eqref{eq:p-Gauss-dist}.
The Besov prior is defined through the wavelet expansion
\eqref{eq:Randomwaveletexpansion}, and the reconstructed signal is $\mathbf{f} =
\mathbf{B}\mathbf{c}$, where $\mathbf{B} = \mathbf{W}_\psi^{T} \mathbf{S}^{-1}$
combines the inverse wavelet transform $\mathbf{W}_\psi^{T}$ and the
diagonal Besov scaling matrix $\mathbf{S}^{-1}$ from \eqref{eq:W_matrix} and
\eqref{eq:S_matrix}, respectively.

Since $\mathbf{c}$ has a smooth, strictly positive density and $\mathbf{B}$ is
invertible, the distribution of $\mathbf{f}$ is absolutely continuous and given by
the pushforward measure 
\[ 
    \mu_f = \mathbf{B}_\# \mu_c, 
    \] 
where $\mu_c$ is the product measure on $\mathbb{R}^n$. While $c_i$ for $i=1,\ldots,n$ are independent,
the components of $\mathbf{f} = \mathbf{B}\mathbf{c}$ are dependent, and the distribution of
$\mathbf{f}$ is no longer product-structured or generalized Gaussian. For $p = 2$, $\mathbf{c}
\sim N(0, \sigma_p^2 I)$ implies $\mathbf{f} \sim N(0, \sigma_p^2
\mathbf{B} \mathbf{B}^T)$, but for $p \ne 2$, no closed-form density
exists.

The distribution of $\mathbf{f}$ remains symmetric about the origin, with marginal tails
still governed by $p$. The covariance matrix of $\mathbf{f}$ is given by 
\[
    \operatorname{cov}(\mathbf{f}) = \mathbf{B} \operatorname{cov}(\mathbf{c}) \mathbf{B}^T = \sigma_p^2 \mathbf{B} \mathbf{B}^T. 
\] 
In Besov priors, where $\mathbf{B} =
\mathbf{W}_\psi^T \mathbf{S}^{-1}$, this becomes 
\begin{equation*}
    \label{eq:covariance_of_Y}
    \operatorname{cov}(\mathbf{f}) = \sigma_p^2 \mathbf{W}_\psi^T \mathbf{S}^{-2} \mathbf{W}_\psi.
\end{equation*}
This covariance matrix is not a multiple of the identity unless $s + d/2 - d/p =
0$ and the wavelet transform is orthonormal, i.e., bi-orthogonal wavelets \emph{or}
non-trivial Besov weights will lead to anisotropic covariance. 

It is the dyadic structure of $\mathbf{S}$ that disrupts the perfect
reconstruction property of the wavelet transform and gives rise to a structured
variation in the standard deviation of the samples, which can be interpreted as
artifacts or patterns in the prior samples.
The structured patterns, illustrated in our numerical experiments, are due to
upsampling in the inverse wavelet transform, variation of the wavelet
filter, and scale-dependent weighting. The resulting uncertainty patterns in
posterior estimates, such as pointwise confidence intervals, arise from this
anisotropic covariance structure.

\bibliography{thebibliography}

@article{Horst2024,
  title   = {Uncertainty Quantification for Linear Inverse Problems with {B}esov Prior: A Randomize-Then-Optimize Method},
  journal = {Statistics and Computing},
  author  = {Horst, Andreas and Maboudi Afkham, Babak and Dong, Yiqiu and Lemvig, Jakob},
  volume  = {35},
  number  = {101},
  year    = {2025},
  doi     = {10.1007/s11222-025-10638-2}
}

@article{Lassas2009,
  title   = {Discretization-invariant {B}ayesian inversion
             and {B}esov space priors},
  journal = {Inverse Problems and Imaging},
  volume  = {3},
  number  = {1},
  pages   = {87-122},
  year    = {2009},
  author  = {Matti Lassas and Eero Saksman and Samuli Siltanen},
  doi     = {10.3934/ipi.2009.3.87}
}

@article{Dashti&Stuart2012,
  title   = {Besov priors for {B}ayesian inverse problems},
  journal = {Inverse Problems and Imaging},
  volume  = {6},
  number  = {2},
  pages   = {183-200},
  year    = {2012},
  author  = {Masoumeh Dashti and Stephen Harris and Andrew Stuart},
  doi     = {10.3934/ipi.2012.6.183}
}

@article{Hoffman2014,
  author  = {Matthew D. Hoffman and Andrew Gelman},
  title   = {The {N}o-{U}-{T}urn {S}ampler: Adaptively setting path lengths in {H}amiltonian {M}onte {C}arlo},
  journal = {Journal of Machine Learning Research},
  year    = {2014},
  volume  = {15},
  number  = {47},
  pages   = {1593--1623}
}

@book{triebel2008,
  title     = {Function Spaces and Wavelets on Domains},
  author    = {Triebel, Hans},
  year      = {2008},
  publisher = {European Mathematical Society},
  address   = {Zürich},
  doi       = {10.4171/019}
}

@book{triebel2006,
  author    = {Triebel, Hans},
  year      = {2006},
  month     = {01},
  title     = {Theory of Function Spaces III},
  doi       = {10.1007/3-7643-7582-5},
  publisher = {Birkhäuser},
  address   = {Basel}
}

@article{suuronen2023bayesian,
  title     = {Bayesian inversion with $\alpha$-stable priors},
  author    = {Suuronen, Jarkko and Soto, Tom{\'a}s and Chada, Neil K and Roininen, Lassi},
  journal   = {Inverse Problems},
  volume    = {39},
  number    = {10},
  pages     = {105007},
  year      = {2023},
  publisher = {IOP Publishing},
  doi       = {10.1088/1361-6420/acf154}
}

@article{suuronen2022cauchy,
  title     = {Cauchy {M}arkov random field priors for {B}ayesian inversion},
  author    = {Suuronen, Jarkko and Chada, Neil K and Roininen, Lassi},
  journal   = {Statistics and computing},
  volume    = {32},
  number    = {2},
  pages     = {33},
  year      = {2022},
  publisher = {Springer},
  doi       = {10.1007/s11222-022-10089-z}
}

@article{uribe2022hybrid,
  title     = {A hybrid {G}ibbs sampler for edge-preserving tomographic reconstruction with uncertain view angles},
  author    = {Uribe, Felipe and Bardsley, Johnathan M and Dong, Yiqiu and Hansen, Per Christian and Riis, Nicolai AB},
  journal   = {SIAM Journal on Uncertainty Quantification},
  volume    = {10},
  number    = {3},
  pages     = {1293--1320},
  year      = {2022},
  publisher = {SIAM},
  doi       = {10.1137/21M1412268}
}

@article{li2022bayesian,
  title     = {Bayesian neural network priors for edge-preserving inversion},
  author    = {Li, Chen and Dunlop, Matthew and Stadler, Georg},
  journal   = {Inverse Problems and Imaging},
  volume    = {16},
  number    = {5},
  pages     = {1229--1254},
  year      = {2022},
  publisher = {Inverse Problems and Imaging},
  doi       = {10.3934/ipi.2022022}
}

@article{uribe2023horseshoe,
  title     = {Horseshoe priors for edge-preserving linear {B}ayesian inversion},
  author    = {Uribe, Felipe and Dong, Yiqiu and Hansen, Per Christian},
  journal   = {SIAM Journal on Scientific Computing},
  volume    = {45},
  number    = {3},
  pages     = {B337--B365},
  year      = {2023},
  publisher = {SIAM},
  doi       = {10.1137/22M1510364}
}

@article{kekkonen2023random,
  title   = {Random tree {B}esov priors--towards fractal imaging},
  author  = {Kekkonen, Hanne and Lassas, Matti and Saksman, Eero and Siltanen, Samuli},
  journal = {Inverse Problems and Imaging},
  volume  = {17},
  number  = {2},
  year    = {2023},
  doi     = {10.3934/ipi.2022059}
}

@article{calvetti2019hierachical,
  title     = {Hierachical {B}ayesian models and sparsity: $\ell$2-magic},
  author    = {Calvetti, Daniela and Somersalo, Erkki and Strang, Alexander},
  journal   = {Inverse Problems},
  volume    = {35},
  number    = {3},
  pages     = {035003},
  year      = {2019},
  publisher = {IOP Publishing},
  doi       = {10.1088/1361-6420/aaf5ab}
}

@article{Daubechies1988Orthonormal,
  author  = {Daubechies, Ingrid},
  title   = {Orthonormal bases of compactly supported wavelets},
  journal = {Communications on Pure and Applied Mathematics},
  volume  = {41},
  number  = {7},
  pages   = {909-996},
  doi     = {https://doi.org/10.1002/cpa.3160410705},
  year    = {1988}
}

@article{Bardsley2014,
  author  = {Bardsley, Johnathan M. and Solonen, Antti and Haario, Heikki and Laine, Marko},
  title   = {Randomize-Then-Optimize: A Method for Sampling from Posterior Distributions in Nonlinear Inverse Problems},
  journal = {SIAM Journal on Scientific Computing},
  volume  = {36},
  number  = {4},
  pages   = {A1895-A1910},
  year    = {2014},
  doi     = {10.1137/140964023}
}

@article{Chung2024,
  author  = {Chung, Julianne and Jiang, Jiahua and Miller, Scot M. and Saibaba, Arvind K.},
  title   = {Hybrid Projection Methods for Solution Decomposition in Large-Scale {B}ayesian Inverse Problems},
  journal = {SIAM Journal on Scientific Computing},
  volume  = {46},
  number  = {2},
  pages   = {S97-S119},
  year    = {2024},
  doi     = {10.1137/22M1502197}
}

@article{Christensen_2024,
  doi       = {10.1088/1361-6420/ad1348},
  year      = {2023},
  month     = {dec},
  publisher = {IOP Publishing},
  volume    = {40},
  number    = {2},
  pages     = {025003},
  author    = {Christensen, Silja L and Riis, Nicolai A B and Pereyra, Marcelo and Jørgensen, Jakob S},
  title     = {A {B}ayesian approach for {CT} reconstruction with defect detection for subsea pipelines},
  journal   = {Inverse Problems}
}

@article{GAO2019,
  title   = {Infimal convolution type regularization of {TGV} and shearlet transform for image restoration},
  journal = {Computer Vision and Image Understanding},
  volume  = {182},
  pages   = {38-49},
  year    = {2019},
  issn    = {1077-3142},
  doi     = {https://doi.org/10.1016/j.cviu.2019.03.002},
  author  = {Yiming Gao and Xiaoping Yang}
}

@article{Chambolle1997,
  author  = {Chambolle, Antonin and Lions, Pierre-Louis },
  title   = {Image recovery via total variation minimization and related problems},
  journal = {Numerische Mathematik},
  volume  = {76},
  number  = {2},
  pages   = {167-188},
  year    = {1997},
  doi     = {10.1007/s002110050258}
}

@article{Starck2005,
  author  = {Starck, Jean-Luc and Elad, Michael and Donoho, David L.},
  journal = {IEEE Transactions on Image Processing},
  title   = {Image decomposition via the combination of sparse representations and a variational approach},
  year    = {2005},
  volume  = {14},
  number  = {10},
  pages   = {1570-1582},
  doi     = {10.1109/TIP.2005.852206}
}

@article{Zhang2021,
  title   = {A customized low-rank prior model for structured cartoon–texture image decomposition},
  journal = {Signal Processing: Image Communication},
  volume  = {96},
  pages   = {116308},
  year    = {2021},
  issn    = {0923-5965},
  doi     = {https://doi.org/10.1016/j.image.2021.116308},
  author  = {Zhiyuan Zhang and Hongjin He}
}

@article{SHI2024,
  title   = {Cartoon-texture guided network for low-light image enhancement},
  journal = {Digital Signal Processing},
  volume  = {144},
  pages   = {104271},
  year    = {2024},
  issn    = {1051-2004},
  doi     = {https://doi.org/10.1016/j.dsp.2023.104271},
  author  = {Baoshun Shi and Chunzi Zhu and Lingyan Li and Huagui Huang}
}

@article{Kaipio2002,
  title   = {Estimating Anomalies from Indirect Observations},
  journal = {Journal of Computational Physics},
  volume  = {181},
  number  = {2},
  pages   = {398-406},
  year    = {2002},
  issn    = {0021-9991},
  doi     = {https://doi.org/10.1006/jcph.2002.7109},
  author  = {Jari P. Kaipio and Erkki Somersalo}
}

@article{Calvetti2011,
  doi       = {10.1088/0266-5611/27/11/115001},
  year      = {2011},
  month     = {oct},
  publisher = {IOP Publishing},
  volume    = {27},
  number    = {11},
  pages     = {115001},
  author    = {Calvetti, Daniela and Homa, Laura and Somersalo, Erkki},
  title     = {Bayesian mixture models for source separation in {MEG}},
  journal   = {Inverse Problems}
}

@inbook{Calvetti2010,
  author    = {Calvetti, Daniela and Somersalo, Erkki},
  publisher = {John Wiley \& Sons, Ltd},
  isbn      = {9780470685853},
  title     = {Subjective Knowledge or Objective Belief? An Oblique Look to {B}ayesian Methods},
  booktitle = {Large‐Scale Inverse Problems and Quantification of Uncertainty},
  chapter   = {3},
  pages     = {33-70},
  doi       = {https://doi.org/10.1002/9780470685853.ch3},
  year      = {2010}
}

@book{Kaipio2005,
  title     = {Statistical and Computational Inverse Problems},
  author    = {Jari P. Kaipio, Erkki Somersalo},
  year      = {2005},
  publisher = {Springer-Verlag},
  address   = {New York},
  doi       = {10.1007/b138659}
}

@inbook{Dashti2017,
  author    = {Dashti, Masoumeh
               and Stuart, Andrew M.},
  title     = {The {B}ayesian Approach to Inverse Problems},
  booktitle = {Handbook of Uncertainty Quantification},
  year      = {2017},
  publisher = {Springer International Publishing},
  address   = {Cham},
  pages     = {311--428},
  doi       = {10.1007/978-3-319-12385-1_7}
}

@article{Senchukova_2024,
  doi       = {10.1088/1361-6420/ad75af},
  year      = {2024},
  month     = {sep},
  publisher = {IOP Publishing},
  volume    = {40},
  number    = {10},
  pages     = {105013},
  author    = {Senchukova, Angelina and Uribe, Felipe and Roininen, Lassi},
  title     = {Bayesian inversion with {S}tudent’s t priors based on {G}aussian scale mixtures},
  journal   = {Inverse Problems}
}

@article{Henderson2024,
  title   = {Sobolev regularity of {G}aussian random fields},
  journal = {Journal of Functional Analysis},
  volume  = {286},
  number  = {3},
  pages   = {110241},
  year    = {2024},
  issn    = {0022-1236},
  doi     = {https://doi.org/10.1016/j.jfa.2023.110241},
  author  = {Iain Henderson}
}

@article{Flock2024,
  title   = {Certified coordinate selection for high-dimensional {B}ayesian inversion with {L}aplace prior},
  journal = {Statistics and Computing},
  volume  = {34},
  number  = {4},
  pages   = {134},
  year    = {2024},
  doi     = {10.1007/s11222-024-10445-1},
  author  = {Flock, Rafael and Dong, Yiqiu and Uribe, Felipe and Zahm, Olivier}
}

@article{Calvetti2020,
  author  = {Calvetti, Daniela and Pragliola, Monica and Somersalo, Erkki},
  title   = {Sparsity Promoting Hybrid Solvers for Hierarchical Bayesian Inverse Problems},
  journal = {SIAM Journal on Scientific Computing},
  volume  = {42},
  number  = {6},
  pages   = {A3761-A3784},
  year    = {2020},
  doi     = {10.1137/20M1326246}
}

@article{Dong_2023,
  doi       = {10.1088/1361-6420/acd851},
  year      = {2023},
  month     = {jun},
  publisher = {IOP Publishing},
  volume    = {39},
  number    = {7},
  pages     = {074001},
  author    = {Dong, Yiqiu and Pragliola, Monica},
  title     = {Inducing sparsity via the horseshoe prior in imaging problems},
  journal   = {Inverse Problems}
}

@article{Waniorek_2023,
  doi       = {10.1088/1361-6420/acad21},
  year      = {2023},
  month     = {feb},
  publisher = {IOP Publishing},
  volume    = {39},
  number    = {2},
  pages     = {024006},
  author    = {Waniorek, Nathan and Calvetti, Daniela and Somersalo, Erkki},
  title     = {Bayesian hierarchical dictionary learning},
  journal   = {Inverse Problems}
}

@article{Glaubitz2024,
  author  = {Glaubitz, Jan and Gelb, Anne},
  title   = {Leveraging Joint Sparsity in Hierarchical {B}ayesian Learning},
  journal = {SIAM/ASA Journal on Uncertainty Quantification},
  volume  = {12},
  number  = {2},
  pages   = {442-472},
  year    = {2024},
  doi     = {10.1137/23M156255X}
}

@article{Laumont2022,
  author  = {Laumont, R\'{e}mi and Bortoli, Valentin De and Almansa, Andr\'{e}s and Delon, Julie and Durmus, Alain and Pereyra, Marcelo},
  title   = {Bayesian Imaging Using Plug \& Play Priors: When {L}angevin Meets {T}weedie},
  journal = {SIAM Journal on Imaging Sciences},
  volume  = {15},
  number  = {2},
  pages   = {701-737},
  year    = {2022},
  doi     = {10.1137/21M1406349}
}

@article{Siltanen2013,
  author  = {H\"{a}m\"{a}l\"{a}inen, Keijo and Kallonen, Aki and Kolehmainen, Ville and Lassas, Matti and Niinim\"{a}ki, Kati and Siltanen, Samuli},
  title   = {Sparse Tomography},
  journal = {SIAM Journal on Scientific Computing},
  volume  = {35},
  number  = {3},
  pages   = {B644-B665},
  year    = {2013},
  doi     = {10.1137/120876277}
}

@article{Kolehmainen_2012,
  doi       = {10.1088/0266-5611/28/2/025005},
  year      = 2012,
  month     = {1},
  publisher = {{IOP} Publishing},
  volume    = {28},
  number    = {2},
  pages     = {025005},
  author    = {Ville Kolehmainen and Matti Lassas and Kati Niinimäki and Samuli Siltanen},
  title     = {Sparsity-promoting {B}ayesian inversion},
  journal   = {Inverse Problems}
}

@article{Schwab2024,
  doi     = {10.1007/s40072-023-00313-w},
  year    = 2024,
  journal = {Stochastics and Partial Differential Equations: Analysis and Computations},
  volume  = {12},
  number  = {3},
  pages   = {1574-1627},
  author  = {Schwab, Christoph and Stein, Andreas},
  title   = {Multilevel {M}onte {C}arlo {FEM} for elliptic {PDE}s with {B}esov random tree priors}
}

@article{Mallat1989,
  author     = {Mallat, Stéphane G.},
  title      = {A Theory for Multiresolution Signal Decomposition: The Wavelet Representation},
  year       = {1989},
  publisher  = {IEEE Computer Society},
  volume     = {11},
  number     = {7},
  issn       = {0162-8828},
  doi        = {10.1109/34.192463},
  journal    = {IEEE Trans. Pattern Anal. Mach. Intell.},
  pages      = {674–693},
  numpages   = {20}
}

@article{Lee2019,
  doi       = {10.21105/joss.01237},
  year      = {2019},
  publisher = {The Open Journal},
  volume    = {4},
  number    = {36},
  pages     = {1237},
  author    = {Gregory R. Lee and Ralf Gommers and Filip Waselewski and Kai Wohlfahrt and Aaron O;Leary},
  title     = {PyWavelets: A {P}ython package for wavelet analysis},
  journal   = {Journal of Open Source Software}
}

@inproceedings{Kutyniok2012,
  author    = {Kutyniok, Gitta
               and Lim, Wang-Q},
  editor    = {Boissonnat, Jean-Daniel
               and Chenin, Patrick
               and Cohen, Albert
               and Gout, Christian
               and Lyche, Tom
               and Mazure, Marie-Laurence
               and Schumaker, Larry},
  title     = {Image Separation Using Wavelets and Shearlets},
  booktitle = {Curves and Surfaces},
  year      = {2012},
  publisher = {Springer Berlin Heidelberg},
  address   = {Berlin, Heidelberg},
  pages     = {416--430}
}

@article{Bobin2007,
  author  = {Bobin, Jerome and Starck, Jean-Luc and Fadili, Jalal M. and Moudden, Yassir and Donoho, David L.},
  journal = {IEEE Transactions on Image Processing},
  title   = {Morphological Component Analysis: An Adaptive Thresholding Strategy},
  year    = {2007},
  volume  = {16},
  number  = {11},
  pages   = {2675-2681},
  doi     = {10.1109/TIP.2007.907073}
}

@article{Gribonval2003,
  author  = {Gribonval, Rémi. and Bacry, Emmanuel},
  journal = {IEEE Transactions on Signal Processing},
  title   = {Harmonic decomposition of audio signals with matching pursuit},
  year    = {2003},
  volume  = {51},
  number  = {1},
  pages   = {101-111},
  doi     = {10.1109/TSP.2002.806592}
}

@article{Liu2019,
  author   = {Liu, Yu and Chen, Xun and Ward, Rabab K. and Wang, Z. Jane},
  journal  = {IEEE Signal Processing Letters},
  title    = {Medical Image Fusion via Convolutional Sparsity Based Morphological Component Analysis},
  year     = {2019},
  volume   = {26},
  number   = {3},
  pages    = {485-489},
  doi      = {10.1109/LSP.2019.2895749}
}

@article{hestenes1952methods,
  title   = {Methods of conjugate gradients for solving linear systems},
  author  = {Hestenes, Magnus R and Stiefel, Eduard and others},
  journal = {Journal of research of the National Bureau of Standards},
  volume  = {49},
  number  = {6},
  pages   = {409--436},
  year    = {1952}
}

@book{mallat1999wavelet,
  title     = {A wavelet tour of signal processing},
  author    = {Mallat, St{\'e}phane},
  year      = {1999},
  publisher = {Elsevier}
}

@article{MR1963681,
  author   = {Donoho, David L. and Elad, Michael},
  title    = {Optimally sparse representation in general (nonorthogonal)
              dictionaries via {$l^1$} minimization},
  journal  = {Proc. Natl. Acad. Sci. USA},
  fjournal = {Proceedings of the National Academy of Sciences of the United
              States of America},
  volume   = {100},
  year     = {2003},
  number   = {5},
  pages    = {2197--2202},
  issn     = {0027-8424,1091-6490},
  mrclass  = {94A29 (94A12)},
  mrnumber = {1963681},
  doi      = {10.1073/pnas.0437847100},
}

@article{Abramovich2002Wavelet,
    author = {Abramovich, F. and Sapatinas, T. and Silverman, B. W.},
    title = {Wavelet Thresholding via A Bayesian Approach},
    journal = {Journal of the Royal Statistical Society Series B: Statistical Methodology},
    volume = {60},
    number = {4},
    pages = {725-749},
    year = {2002},
    month = {01},
    issn = {1369-7412},
    doi = {10.1111/1467-9868.00151},
}

@article{Donoho1994Ideal,
    author = {Donoho, David L and Johnstone, Iain M},
    title = {Ideal spatial adaptation by wavelet shrinkage},
    journal = {Biometrika},
    volume = {81},
    number = {3},
    pages = {425-455},
    year = {1994},
    month = {09},
    issn = {0006-3444},
    doi = {10.1093/biomet/81.3.425},
}

@article{Donoho1995Adapting,
author = {David L. Donoho and Iain M. Johnstone},
title = {Adapting to Unknown Smoothness via Wavelet Shrinkage},
journal = {Journal of the American Statistical Association},
volume = {90},
number = {432},
pages = {1200--1224},
year = {1995},
publisher = {ASA Website},
doi = {10.1080/01621459.1995.10476626}
}

@article{Donoho1998Minimax,
author = {David L. Donoho and Iain M. Johnstone},
title = {{Minimax estimation via wavelet shrinkage}},
volume = {26},
journal = {The Annals of Statistics},
number = {3},
publisher = {Institute of Mathematical Statistics},
pages = {879 -- 921},
year = {1998},
doi = {10.1214/aos/1024691081},
}

@article{Agapiou2024Adaptive,
author = {Sergios Agapiou and Aimilia Savva},
title = {{Adaptive inference over Besov spaces in the white noise model using p-exponential priors}},
volume = {30},
journal = {Bernoulli},
number = {3},
publisher = {Bernoulli Society for Mathematical Statistics and Probability},
pages = {2275 -- 2300},
year = {2024},
doi = {10.3150/23-BEJ1673},
}

@article{Giordano2023Besov,
author = {Matteo Giordano},
title = {{Besov-Laplace priors in density estimation: optimal posterior contraction rates and adaptation}},
volume = {17},
journal = {Electronic Journal of Statistics},
number = {2},
publisher = {Institute of Mathematical Statistics and Bernoulli Society},
pages = {2210 -- 2249},
year = {2023},
doi = {10.1214/23-EJS2161},
}

@article{Cotter2013MCMC,
author = {S. L. Cotter and G. O. Roberts and A. M. Stuart and D. White},
title = {{MCMC Methods for Functions: Modifying Old Algorithms to Make Them Faster}},
volume = {28},
journal = {Statistical Science},
number = {3},
publisher = {Institute of Mathematical Statistics},
pages = {424 -- 446},
year = {2013},
doi = {10.1214/13-STS421},
}
\end{document}